\documentclass[aps,prb,twocolumn,floatfix]{revtex4}
\usepackage{color,graphicx}
\usepackage{epsfig}
\usepackage{subfigure}
\usepackage{amsmath,amsfonts,amssymb,amsthm,verbatim}
\usepackage{tabularx}
\usepackage{bm}

\def\be{\begin{equation}}       \def\ee{\end{equation}}
\def\bea{\begin{eqnarray}}      \def\eea{\end{eqnarray}}

\def\be{\begin{equation}}
\def\ee{\end{equation}}
\def\ba{\begin{eqnarray}}
\def\ea{\end{eqnarray}}
\def\Cal{\mathcal}
\def\ie{{\it i.e.}}

\begin{document}
\title{Quasiparticle scattering in two dimensional helical liquid}
\author{Xiaoting Zhou}
\affiliation{Department of Physics, Purdue University, West Lafayette, IN 47907}
\author{Chen Fang}
\affiliation{Department of Physics, Purdue University, West Lafayette, IN 47907}
\author{Wei-Feng Tsai}
\affiliation{Department of Physics, Purdue University, West Lafayette, IN 47907}
\author{JiangPing Hu}
\affiliation{Department of Physics, Purdue University, West Lafayette, IN 47907}
\date{\today}
\newcommand{\br}{\mathbf{r}}
\newcommand{\brprime}{{\mathbf{r}^\prime}}
\newcommand{\bk}{\mathbf{k}}
\newcommand{\bq}{\mathbf{q}}

\begin{abstract}
%Spotlight of our work.
We study the quasiparticle interference (QPI) patterns caused by
scattering  off nonmagnetic, magnetic point impurities, and edge
impurities, separately, in a two dimensional helical liquid, which
describes the   surface states of a topological insulator. The
unique features associated with hexagonal warping effects are
identified in the QPI patterns of charge density with nonmagnetic
impurities and spin density with magnetic impurities. The symmetry
properties of the QPI patterns can be used to determine the symmetry
of microscopic models.  The Friedel oscillation is calculated for
edge impurities and the decay of the oscillation is not universal,
strongly depending on Fermi energy. Some discrepancies between our
theoretical results and current experimental observations are
discussed.
\end{abstract}
\maketitle

\section{Introduction}
Several recent theoretical\cite{fu07a,qi08,zhangiop} and
experimental\cite{hsieh08,hsieh09a,hsieh09b,chen09,xia09,hor09}
works have focused on a new quantum state of matter, {\it
topological insulators} in three dimensions, which exhibit bulk
insulating gaps (mainly of spin-orbit origin) while possess {\it
time-reversal symmetry} protected gapless surface states. One of
intriguing properties in this new quantum state comes from those
``protected'' surface states, which provide a lab-realizable
condensed-matter analog of two dimensional, massless Dirac theory
with ``odd'' number of species (Dirac cones), in the surface
Brillouin zone (SBZ)\cite{fu07a,fu07b}. The charge carriers on the
surfaces here, the so-called (spin) {\it helical} Dirac
fermions\cite{wu06,hsieh09b}, behave like relativistic particles
with a spin locked to its momentum leading to the breakdown of the
spin $SU(2)$ rotational symmetry. This feature is sharply in
contrast to graphene, where the system not only possesses an even
number of Dirac cones in its spectrum, but the role of the
``locked'' spin is also replaced by a pseudo-spin (sublattice
symmetry) and hence each Dirac cone still has two-fold spin
degeneracy\cite{geim07}.
%Fundamentally, it is just the ``boundary'' nature of helical Dirac fermion systems that the odd number of particle species can exist without violating the fermion doubling rule in 2D systems\cite{nielsen81}.
%[*any exciting consequence?]

As a useful surface probe, recent angle-resolved photoemission
spectroscopy (ARPES) experiments successfully demonstrated the
surface band structures with odd number of Dirac
cones\cite{chen09,hsieh08} as well as the corresponding spin
helical structures near a Dirac
point\cite{xia09,hsieh09a,hsieh09b}. Although the confirmed nature
of the bands by ARPES suggests the quantum state to be
topologically insulating, the quest for new quantum phenomena
uniquely associated with such topology-protected surface states
remains urgent and necessary. The usual way in solid state physics
to explore the nontrivial electronic properties of helical Dirac
fermion systems would be the transport measurement on the surface
of a topological insulator\cite{fu07b}. However, such a
measurement may not be practically straightforward, since (i)
tuning the system to the topological transport regime where the
charge density vanishes is tricky, and (ii) the presence of the
n-type doping from vacancy (or anti-site defects) as well as the
fact that the surface states surround the sample make the results
difficult to be distinguished from the bulk and surface
contributions\cite{chen09,peng09}.

Alternatively, the quasiparticle interference (QPI) caused by
scattering off impurities on a surface can provide a way of
revealing the topological nature of the surface
states\cite{yazdani2009,alpichshev09,zhang09,gomes09}. The concept
of QPI is elementary in quantum mechanics. For instance, due to
impurity (elastic) scattering, the interference between the incoming
and outgoing waves with momenta $\bk_i$ and $\bk_f$, respectively,
can give rise to an amplitude modulation in the local density of
states (LDOS) at wavevector $\mathbf{q}=\bk_f-\bk_i$. Such kind of
interference pattern can be observed in Fourier transform scanning
tunneling spectroscopy (FT-STS) nowadays and it has been proved
useful in determining the pairing nature of high-$T_c$
cuprates\cite{hoffman02}. By measuring the QPI patterns and
analyzing them through a convolution of  ARPES data together with a
spin-dependent scattering matrix element, Roushan and et
al\cite{yazdani2009} were able to demonstrate the absence of
backscattering in the topological surface states of
$Bi_{1-x}Sb_{x'}$, a key property of helical spin liquid.

Most recently, based on symmetry analysis,  a new hexagonal warping
term, which is absent in $Bi_{1-x}Sb_{x'}$, is suggested by Liang
Fu\cite{fu09} to explain the evolution of the Fermi surface of the
effective 2D helical Dirac model describing the surface band
structure of a family of 3D topological insulators, $Bi_2X_3$ (X=Se
or Te). As measured in ARPES experiments,  the shape of the Fermi
surface (FS) evolves gradually from a hexagram, a hexagon, to a
circle of shrinking volume, and finally meets at the Dirac point
when lowering the Fermi energy. The new term leads to strong density
variation around Fermi surface and also modifies the spin helical
configuration. As a result, the existence of the new term can
strongly modify the QPI. In other words, the QPI can provide a
direct evidence to justify the model.

In this paper we systematically investigate the interference effects
of a point-impurity  and an edge-impurity scattering, respectively,
on the LDOS in a 2D helical Dirac fermion system. We use $T$-matrix
approach to calculate QPI spectra at  a few representative energies,
for emphasizing the effects of the hexagonal warping term, in the
presence of a nonmagnetic/magnetic impurity. We also investigate an
edge impurity by using a method generalized from 1D scattering
problems with a potential barrier. Several profound features are
found in this study. In a nutshell, we observe: (i) the backward
scattering by nonmagnetic point impurities is topologically
suppressed, just as what has been shown in \cite{yazdani2009} with a
simpler empirical analysis,  and the dominant interference pattern
becomes that of spatial period $2\pi/|\bq_{35}|$ when going away
from the Dirac regime (see Fig.~\ref{fig:CCE_EW} for the definition
of $\bq_{35}$); (ii) In the presence of magnetic impurity,   the QPI
of charge density  is very weak while that of spin density becomes
strong. Near the Dirac regime, spin moments of fermions are flipped
when scattering wave vector crosses over $|\bq|=2|\bk_F|$, as
demonstrated in the ($z$-component) spin LDOS [see
Fig.~\ref{fig:dmagz_QPI} (b)]; (iii) the mirror symmetries of the
spin LDOS in the presence of in-plane magnetic impurity with spin
polarization fixed along $x$ and $y$ directions   can be used to
determine the symmetry of microscopic models and to verify the
presence or absence of the warping term; (iv)  In the case of 1D
edge impurities, the Friedel oscillation has no universal decaying
function. Depending on Fermi surface energy, we show that the
oscillation decays as $1/\sqrt{|x|}$ if the FS shape is dominated by
the warping term, and as $|x|^{-3/2}$ if the warping term is
negligible.
  These special quantum phenomena, sharply in contrast to conventional
  metals, are mainly associated with the 2D helical liquid.

\begin{figure}[th]
\begin{center}
\includegraphics[scale=0.5]{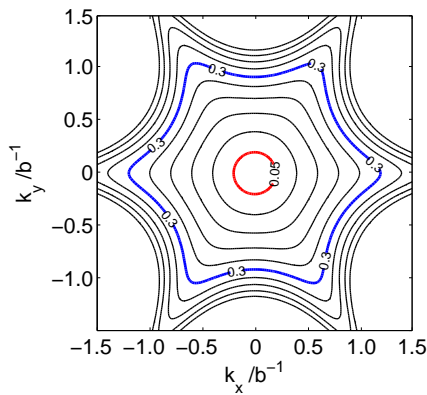}
\caption{Contours of constant energy and the evolution of FS.}
\label{fig:cce}
\end{center}
\end{figure}

\section{The model and $T$-matrix formalism}
\begin{figure}[tbh]
\begin{center}
\includegraphics[scale=0.45]{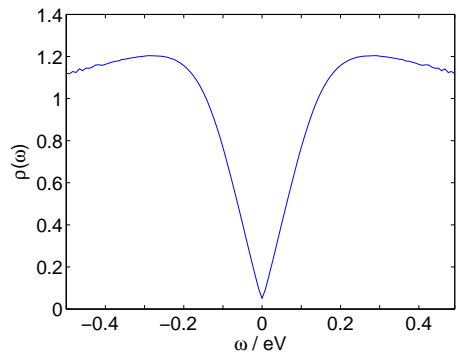}
\caption{Density of states based on the model in Eq.~(\ref{eq:model}).}
\label{fig:dos}
\end{center}
\end{figure}

We now briefly introduce our used formalism below. The explicit
model we study here is written as \be
H(\bk)=v(k_x\sigma_y-k_y\sigma_x)+\frac{k^2}{2m^*}+\frac{\lambda}{2}(k^3_+
+k^3_-)\sigma_z, \label{eq:model} \ee where $k_{\pm}\equiv k_y \pm i
k_x$. $v$ and $\lambda$ denote Fermi velocity and  hexagonal warping
parameter, respectively. The Pauli matrices, $\sigma_i$, act on spin
space of fermionic quasiparticles. The form of $H(\bk)$ is suitable
for describing the [111] surface band structure near $\Gamma$ point
in SBZ of a 3D topological insulator Bi$_2$X$_3$, and is fixed under
general symmetry considerations, namely, time reversal and $C_{3v}$
symmetries\cite{fu09}. Notice that we have chosen $x$ direction to
be along $\Gamma M$ in SBZ. The $k$-linear term,
$H_0=v(k_x\sigma_y-k_y\sigma_x)$, describes an isotropic 2D helical
Dirac fermions, and the $k$-square term causes particle-hole
asymmetry. More importantly, the $k$-cube warping term,
$H_w=\frac{\lambda}{2}(k^3_+ +k^3_-)\sigma_z$, leads to hexagonal
distortion of the Fermi surface. The resulting two energy bands now
touch at the Dirac point (\ie, $\Gamma$ point in SBZ) with
dispersion relation, \be
\epsilon_{\pm}(\bk)=\frac{k^2}{2m^*}\pm\sqrt{v^2k^2+\frac{{\lambda}^2}{4}(k_+^3+k_-^3)^2}.
\ee Defining the characteristic length scale
$b\equiv\sqrt{\lambda/v}$ and energy $E^*\equiv v/b$ introduced by
the hexagonal warping parameter, we draw the contours of constant
energy (CCE) in momentum space in units of $1/b$ and single-particle
density of states (DOS) of $H(\bk)$, respectively, in
Figs.~\ref{fig:cce} and \ref{fig:dos}.

In the numerical evaluation, we have taken $b\equiv 1$
%($\approx 2.5a$ in experiments,  $a=$ lattice constant)
, $v=0.25$, and $\lambda=0.25$ such that the Fermi surface in 0.67\%
Sn-doped Bi$_2$Te$_3$ can  be qualitatively reproduced, where the
measured $v=2.55eV\cdot\r{A}$ and $E_F=1.2E^*\approx$0.3eV. Unless
otherwise stated, we will assume particle-hole symmetry, \ie,
$m^*\rightarrow \infty$. As shown in Figs.~\ref{fig:cce} and
\ref{fig:dos}, when $\omega\ll 0.2$ the DOS is almost linear in
$\omega$ with more circular FS, while when $\omega\gg 0.2$ the DOS
behaves like $\omega^{-1/3}$ with hexagram-like FS.

\begin{figure}[tbh]
\begin{center}
\includegraphics[scale=0.5]{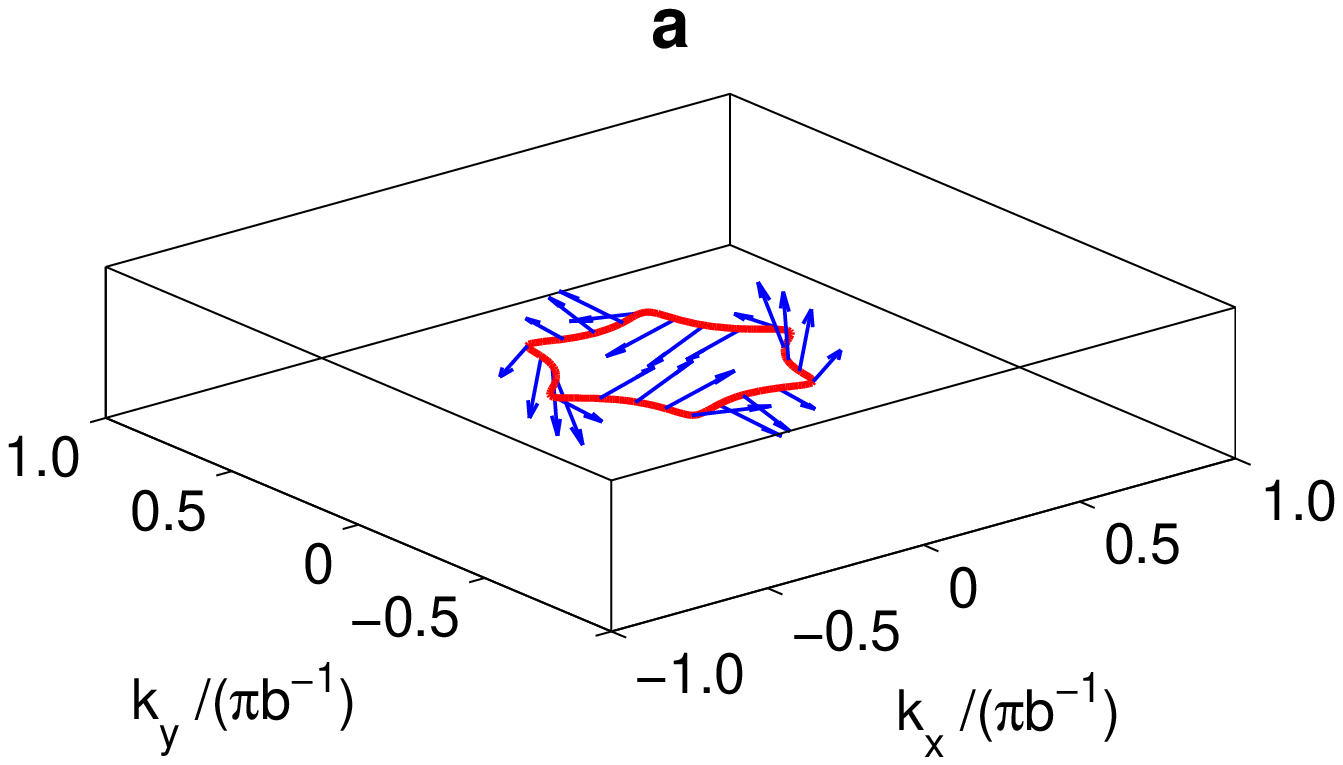}
\includegraphics[scale=0.5]{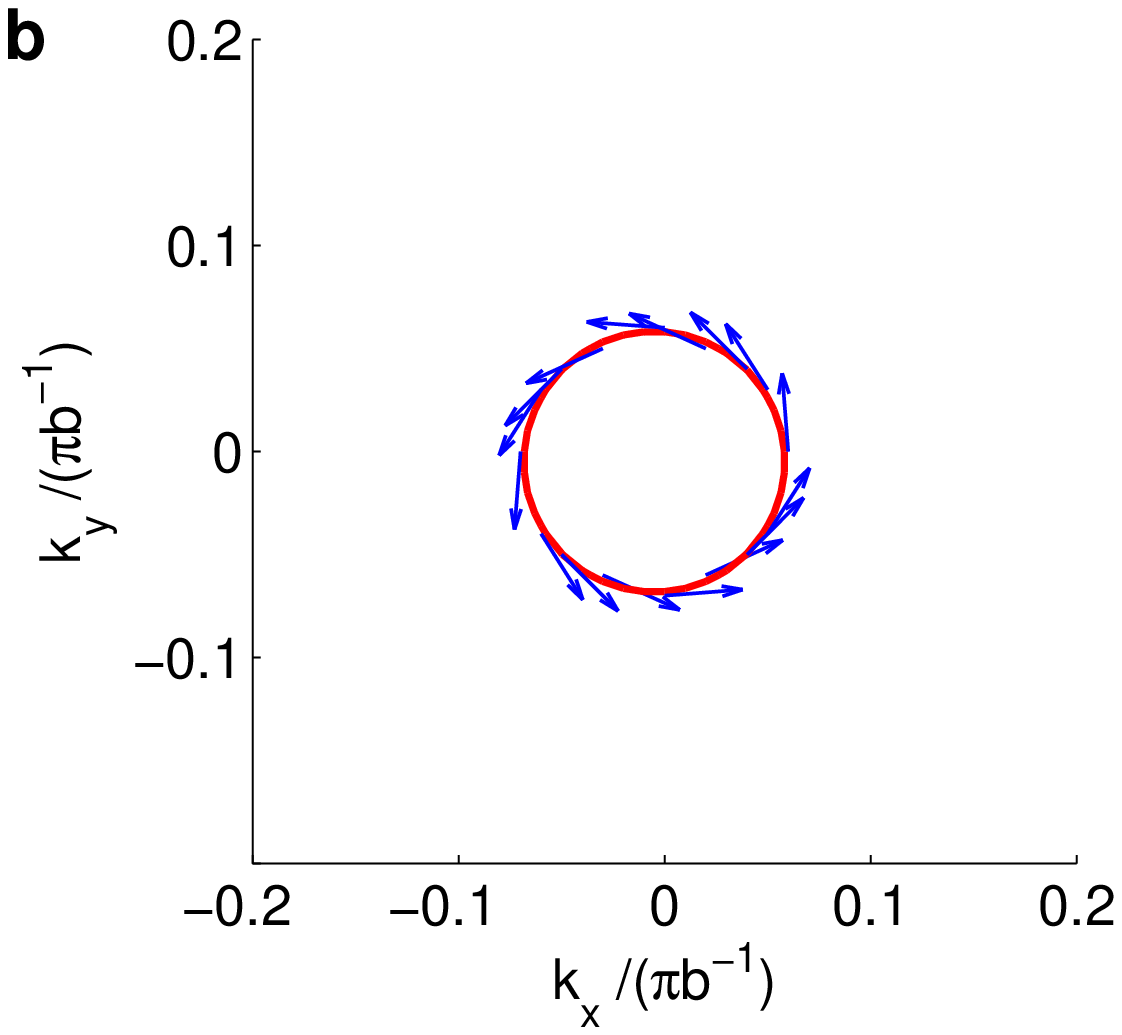}
\caption{Spin textures around the Fermi surface at $\omega=0.3$eV in
(a) and at $\omega=0.05$eV (b). \label{fig:sFS}}
\end{center}
\end{figure}

In addition to the CCE, we also present the spin-resolved FS with
two  representative energies used through out this paper,
$E_D=$0.05eV ($0.2E^*$) and $E_W=$0.3eV ($1.2E^*$) in
Fig.~\ref{fig:sFS}. They clearly demonstrate the ``spin-helical''
nature of the 2D fermions, which is indeed essential when analyzing
the QPI spectra later. In particular, as $\omega=E_W$, non-vanishing
spin moments along z direction (out of surface plane) are present
mainly due to  $\sigma_z$ in the warping term, which is directly
proportional to electron's spin. Notice that the spin moment must be
in-plane along $\Gamma M$ (\ie, at each sharp vertex of the FS),
which is a consequence of the odd parity of $\sigma_z$ under the
mirror operation $y\rightarrow -y$.

Next, we consider the quasiparticle scattering problem within the
$T$-matrix approach\cite{morr03}. For a general $N$-impurity
problem, the impurity-induced electronic Green's function is given
by \be \delta G(\br,\br^\prime,\omega) =
\sum_{i,j=1}^{N}G_0(\br,\br_i,\omega)
T(\br_i,\br_j,\omega)G_0(\br_j,\br^\prime,\omega),
\label{eq:rspaceT} \ee where the $T$-matrix obeys the
Bethe-Salpeter equation \be T(\br_i,\br_j,\omega) =
V_{\br_i}\delta_{\br_i,\br_j}+V_{\br_i}\sum_{k=1}^{N}
G_0(\br_i,\br_k,\omega)T(\br_k,\br_j,\omega), \label{eq:BSeq} \ee
and the Green's function (in momentum space) of the clean system
is \be G_0(\bk,\omega) = [\omega+i\eta-H(\bk)]^{-1}. \label{eq:GF}
\ee

In the case of a single point nonmagnetic (magnetic) impurity
located at  the origin, the scattering potential is simply
$V_{\br}=\delta_{\br,0}V_{NI}\sigma_0$
$(\delta_{\br,0}V_{MI}\vec{\sigma})$, where $\sigma_0$ is a $2\times
2$ identity matrix. Taking the advantages of the translational
symmetry of the clean system and momentum independence of the
scattering potential (for instance,
$V_{\bk,\bk^\prime}=V_{NI}\sigma_0/N\equiv\hat{V}$ in the
nonmagnetic case), one can simplify the formula as \be
T(\omega)=[1-\hat{V}\int_{\epsilon_+(\bk)<\Lambda}\frac{d^2
k}{(2\pi)^2} G_0(\bk,\omega)]^{-1}\hat{V}, \ee and hence around the
impurity, spatial oscillations of the local density of states are
induced. To see the interference effects due to impurity scattering,
it is more convenient to compute the Fourier-transformed (induced)
local density of states (FT-LDOS), \ba
\int d^2 r e^{i\bq\cdot\br}\delta\rho(\br,\omega) &\sim& \delta\rho(\bq,\omega) \nonumber \\
&=& \frac{i}{2\pi}\int_{\epsilon_+(\bk)<\Lambda}\frac{d^2
k}{(2\pi)^2}g(\bk,\bq,\omega), \ea where
$g(\bk,\bq,\omega)=\sum_{i=1}^{2}[\delta
G_{ii}(\bk,\bk+\bq,\omega)-\delta G^*_{ii}(\bk+\bq,\bk,\omega)]$.
  In general, $\rho(\bq,\omega)$ is  a complex number. If we
separately define the symmetric and antisymmetric parts of the
LDOS as
$\rho^S(x,y,\omega)=[\rho(x,y,\omega)+\rho(-x,-y,\omega)]/2$ and
$\rho^A(x,y,\omega)=[\rho(x,y,\omega)-\rho(-x,-y,\omega)]/2$, the
real and imaginary parts of $\rho(\bq,\omega)$ simply describe the
symmetric and antisymmetric parts of the LDOS respectively. In the
following discussion of the effects of non-magnetic impurities,
since the real part is at least two orders of magnitude larger
than the imaginary part, we focus on the former. In our
calculation, we have introduced an energy cutoff $\Lambda=4 E^*$
when integrating over momentum. Our main results do not
sensitively depend on the chosen $\Lambda$ as long as $\Lambda$ is
much greater than the impurity scattering strength. Moreover, the
spin-resolved FT-LDOS can be obtained if we separate each
component $i$ when evaluating function $g(\bk,\bq,\omega)$, \ie,
$i=1$ for spin-up and $i=2$ for spin-down.

In principle, for the case of an edge-impurity scattering, one can
use Eqs.~(\ref{eq:rspaceT})-(\ref{eq:GF}) to compute the LDOS from
$\delta\rho(\br,\omega)$=-Im$\sum_{i}\delta
G_{ii}(\br,\br,\omega)/\pi$ in a straightforward manner. However,
it is more convenient, without loss of generality, to treat this
scattering problem by using an analogy of the elementary
scattering problem with a barrier potential in one dimension,
which is directly based on the wave function point of view. Our
method is briefly sketched in section III C.

%In the case of a (nonmagnetic) line impurity scattering, we may set $V_{\br}=\delta_{x,0}V_{LI}\sigma_0$ without loss of generality. In principle, one can directly use Eqs.~(\ref{eq:rspaceT})-(\ref{eq:GF}) to compute the LDOS from $\delta\rho(\br,\omega)$=-Im$\sum_{i}\delta G_{ii}(\br,\br,\omega)/\pi$. However, by realizing the fact that for a line impurity scattering the $y$ component of the incoming and outgoing momenta should be conserved, one may obtain LDOS by a modified formula similar to the point-impurity case as follows,
%\be
%T_{k_y}(\omega)=[1-\hat{V}\int_{\epsilon_+(\bk)<\Lambda}\frac{dk_x}{2\pi} G_0(k_x,k_y,\omega)]^{-1}\hat{V},
%\ee
%and henceforth,
%\be
%\delta\rho(q_x,\omega)=\frac{i}{2\pi}\int_{\epsilon_+(\bk)<\Lambda}\frac{d^2 k}{(2\pi)^2}g(\bk,q_x,\omega)
%\ee
%with corresponding $g(\bk,q_x,\omega)=\sum_{i=1}^{2}[\delta G_{ii}(\bk,\bk+q_x \hat{x},\omega)-\delta G^*_{ii}(\bk+q_x \hat{x},\bk,\omega)]$ and $\delta G(\bk,\bk+q_x \hat{x},\omega)=G_0(\bk,\omega)T_{k_y}(\omega)G_0(\bk+q_x \hat{x},\omega)$.

\section{Numerical results}
We compute the induced LDOS at selected $\omega$,
$\delta\rho(\bq,\omega)$, for the  nonmagneic/magnetic impurity
case, and, $\rho(q_x,\omega)$, for the edge impurity case. Our
numerical results are reported for a representative potential
scattering strength, $V_{NI}=V_{MI}=V_0=$0.05eV. The chosen imaginary
part of the energy $\eta=$10meV has been checked to be insensitive
to the observed main features. Also, in our analysis a $400 \times
400$ momentum grid is used in $(-\pi,\pi)\times(-\pi,\pi)$ $k$ space
and $200$ discrete points are displayed within $(-\pi,\pi)$ along
each direction in $q$ space. Note that the relevant range of SBZ in
experiments would correspond to about 5.5 times larger than $2\pi$.

\begin{figure}[tbh]
\begin{center}
\includegraphics[scale=0.65]{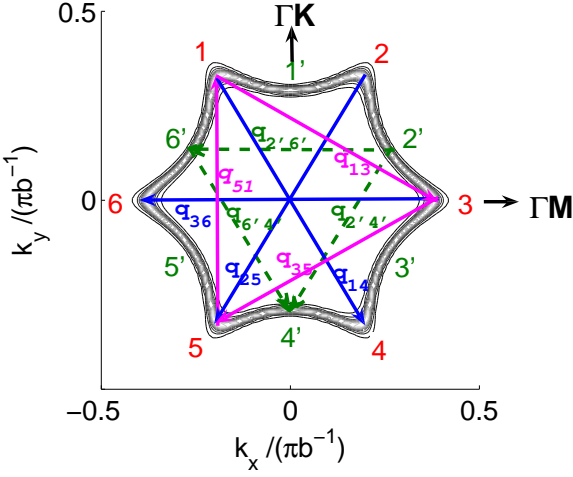}
\caption{The spectral function $\Cal{A}(\bk,\omega)$ at
$\omega=0.3$eV with three most possible scattering wave vectors.
Note the wave vector is in units of $\pi b^{-1}$ and brighter
region corresponds to higher spectral weight.} \label{fig:CCE_EW}
\end{center}
\end{figure}

\subsection{Nonmagnetic point impurity}
We first consider the interference patterns in a 2D helical liquid
with  a nonmagnetic point impurity. Starting with
$\omega=E_W=$0.3eV far away from the Dirac point ($\omega=0$), the
shape of the FS is now like a hexagram. This is just the energy
range where experiment may achieve without subtle chemical tuning
near the surface of a 3D topological insulator. As we will see
later, such energy range indeed provide a better chance to reveal
the topological nature of the helical Fermion system. In
Fig.~\ref{fig:CCE_EW}, the spectral function,
$\Cal{A}(\bk,\omega)=-\frac{1}{\pi}\text{Im}[\text{Tr}G_0(\bk,\omega)]$
at $\omega=0.3$eV, are plotted with scattering vectors on top,
which are expected to associate with high joint DOS on a
constant-energy contour.

\begin{figure}[tbh]
\begin{center}
\includegraphics[scale=0.45]{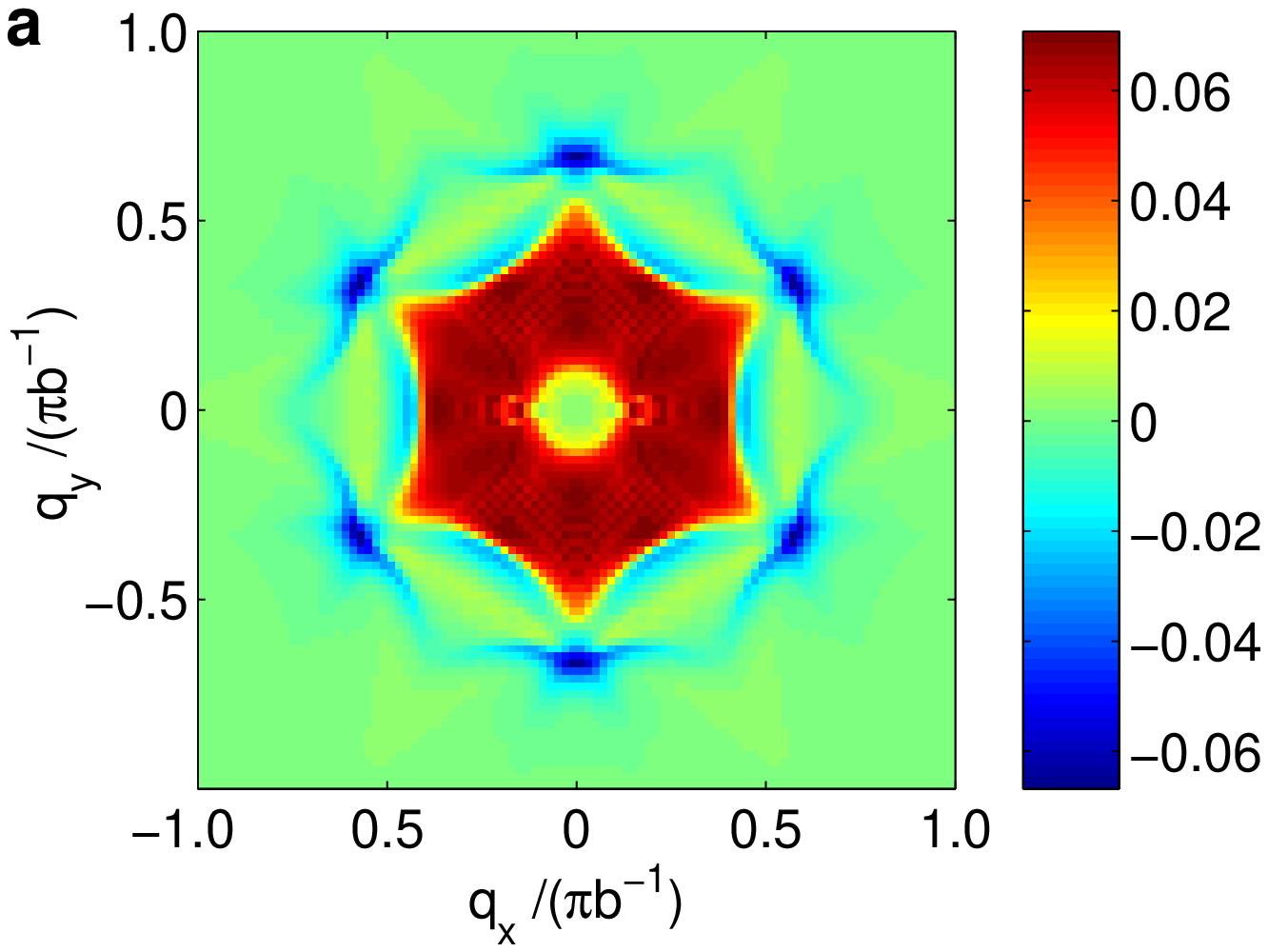}
\includegraphics[scale=0.45]{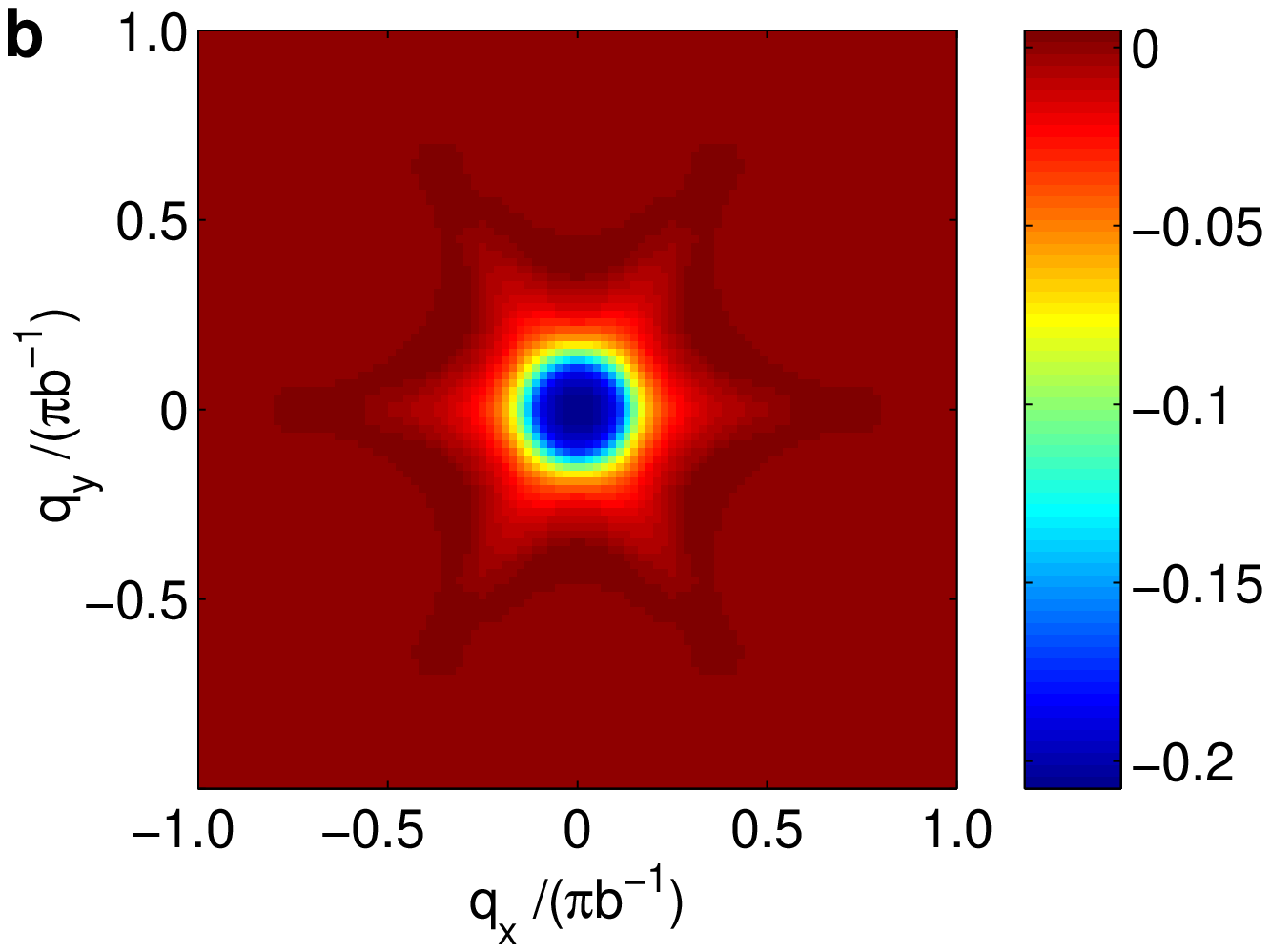}
\caption{The real part of the Fourier transform of local density of
states in the case of single nonmagnetic point impurity at (a)
$\omega=0.3eV$, and (b) $\omega=0.05eV$.} \label{fig:nonmag_QPI}
\end{center}
\end{figure}

As shown in Fig.~\ref{fig:nonmag_QPI} (a), the interference pattern
includes six sharp peaks along $\Gamma K$ outside a complicated,
hexagon-shaped pattern centered at $\Gamma$ and other six weaker
peaks along $\Gamma M$ slightly inside the hexagon. These two sets
of peaks simply correspond to
$(\pm\bq_{13},\pm\bq_{35},\pm\bq_{51})$ and
$(\pm\bq_{12},\pm\bq_{23},\pm\bq_{34})$, respectively, as indicated
in Fig.~\ref{fig:CCE_EW}. However, the most prominent feature we
observed here is that those expected peaks, which correspond to the
$(\pm\bq_{14},\pm\bq_{25},\pm\bq_{36})$, are entirely absent. This
apparent puzzle can be understood by the absence of backscattering
between two time reversal connected partners, as shown in
\cite{yazdani2009}. Suppose in our scattering problem,
$|\bk,\uparrow\rangle$ is the incoming state, while its
time-reversal partner,
$|-\bk,\downarrow\rangle\propto\Cal{T}|\bk,\uparrow\rangle$, is the
outgoing state. $\Cal{T}$ is the time-reversal operator with the
property $\Cal{T}^2=-1$. For any time-reversal invariant and
hermitian operator $\hat{V}$ (such as our nonmagnetic scattering
potential), we have \ba
\langle-\bk,\downarrow|\hat{V}|\bk,\uparrow\rangle &=& \langle
\Cal{T}(\bk,\uparrow)|\hat{V}(\bk,\uparrow)\rangle=\langle\Cal{T}\hat{V}
(\bk,\uparrow)|\Cal{T}^2(\bk,\uparrow)\rangle \nonumber \\
&=& -\langle\bk,\uparrow|\Cal{T}\hat{V}|\bk,\uparrow\rangle^* =-\langle\bk,\uparrow|\hat{V}\Cal{T}|\bk,\uparrow\rangle^* \nonumber \\
&=& -\langle\bk,\uparrow|\hat{V}|-\bk,\downarrow\rangle^* =
-\langle -\bk,\downarrow|\hat{V}^\dagger|\bk,\uparrow\rangle \nonumber \\
&=& -\langle -\bk,\downarrow|\hat{V}|\bk,\uparrow\rangle =0.
\label{eq:forbidden} \ea In other words, the backward scattering
between time-reversal partners is not allowed.  This naturally
explains the absence of the interference peaks, corresponding to
$\bq_{36}$ (and of the same type). Such a behavior sharply
distinguishes the 2D helical Fermion system from a conventional
metal. In addition, it might be worth mentioning here that the
angles of our observed interference peaks, $\bq_{35}$, appear
different from the experiment done by Zhang {\it et
al.}\cite{zhang09}, where there exhibits six peaks along $\Gamma M$,
instead of $\Gamma K$ as displayed in Fig.~\ref{fig:nonmag_QPI} (a).
We would like to postpone this issue to the discussion section.

\begin{figure}[tbh]
\begin{center}
\includegraphics[scale=0.45]{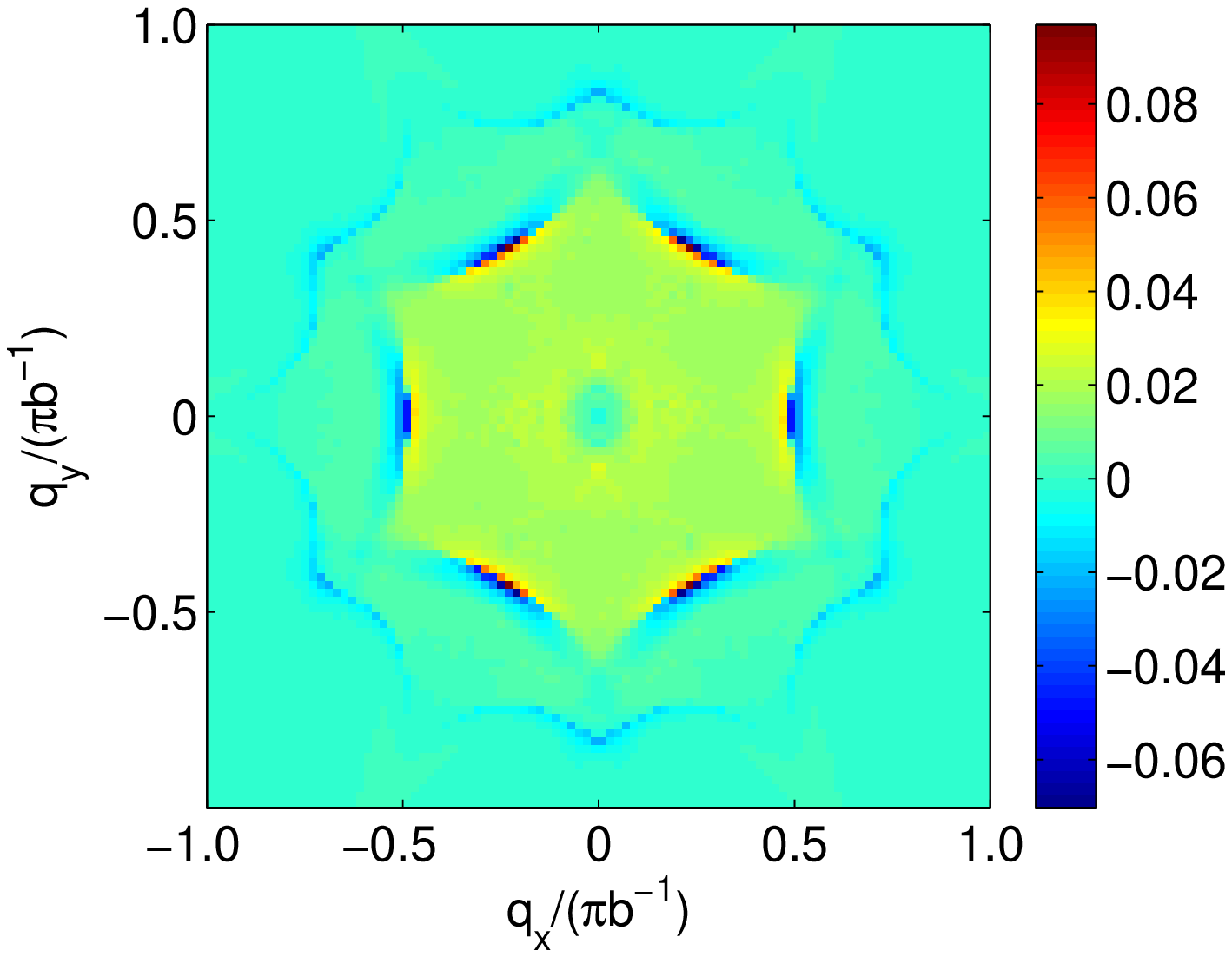}
\caption{The real part of the Fourier transform of local density of
states in the case of single nonmagnetic point impurity at
$\omega=0.375eV$. \label{fig:nonmag_0375}}
\end{center}
\end{figure}

When further increasing the Fermi level, the vertices become sharper
and the joint  DOS at fixed $\bq_{35}$, however, is suppressed. As a
result, the six peaks seen in Fig.~\ref{fig:nonmag_QPI}(a) diminish
and the replaced feature turns out to be the other six peaks
\textbf{at fixed $\bq'$}, corresponding to the scattering vectors
connecting between second neighbor of the convex parts of the FS
(see Fig.~\ref{fig:nonmag_0375}), which were observed in recent
experiments\cite{zhang09}. On the other hand, when the Fermi level
gets closer to the Dirac point, for instance, $\omega=$0.05eV, the
interference pattern becomes almost isotropic with obvious stronger
weight within a circular region, as shown in
Fig.~\ref{fig:nonmag_QPI} (b). The size of the region can be
estimated to be a disk with twice longer radius of the corresponding
circular FS of the system. This is basically consistent with our CCE
picture (see Fig.~\ref{fig:cce}), where no finite, specific $\bq$
vectors can be picked out when $\omega$ approaches to the Dirac
point.

\subsection{Classical magnetic point impurity}
%It is interesting to
Next, we study the QPI induced by a time-reversal symmetry breaker,
a magnetic  impurity\cite{liu09}.  We focus on the effects of a
classical magnetic impurity so that the Kondo physics is ignored. In
the following, after describing general features of the QPI with a
magnetic impurity,
%the results for a general magnetic impurity,
we will   discuss the cases separately when the impurity moment is
fixed along $x$, $y$, and $z$ directions.
\begin{figure}[tbh]
\begin{center}
\includegraphics[scale=0.4]{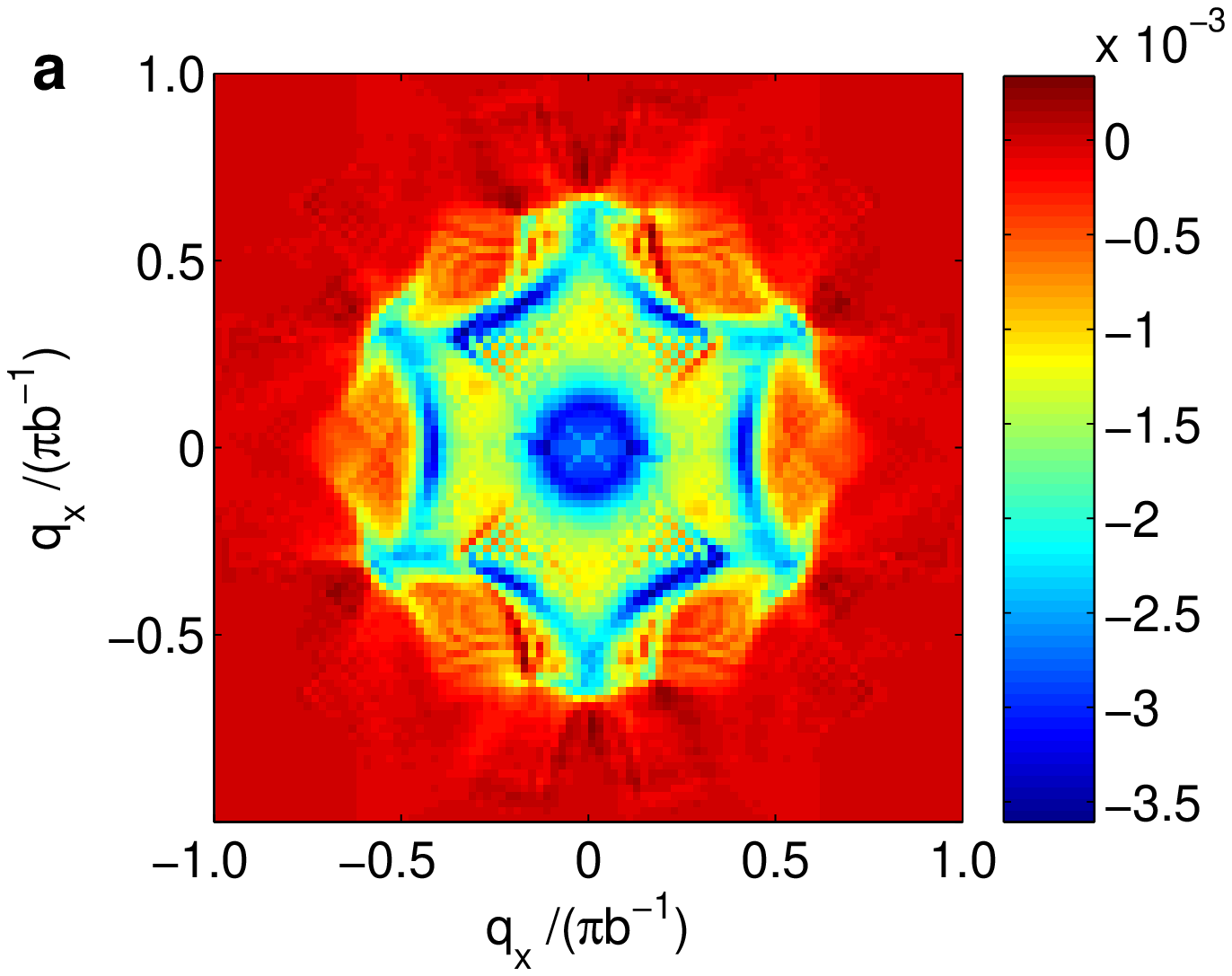}
\includegraphics[scale=0.4]{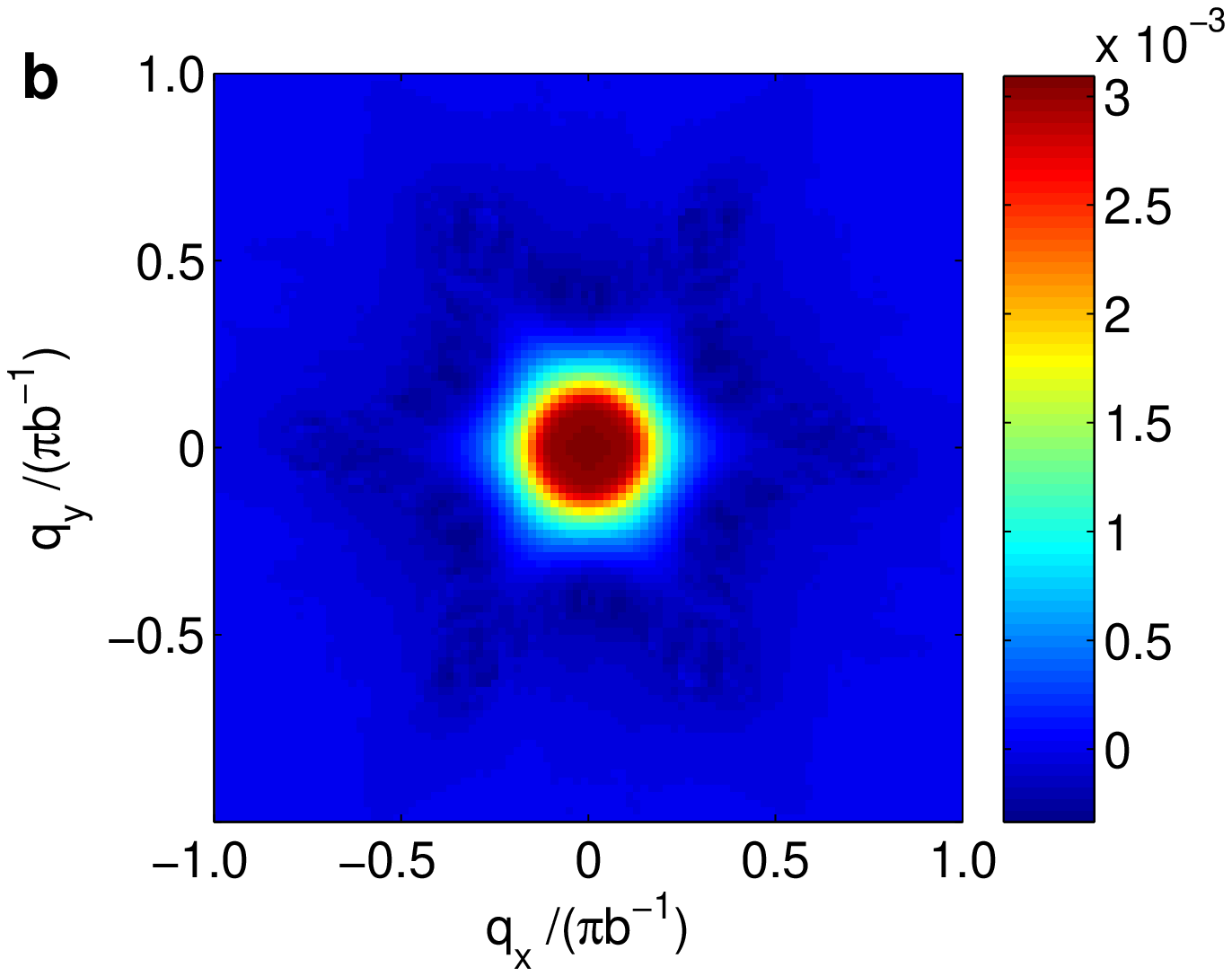}
\caption{The real part of the Fourier transform of charge local
density of states in the case of single magnetic point impurity
with its spin polarized along the z-axis at (a) $\omega=0.3eV$,
and (b) $\omega=0.05eV$.} \label{fig:tmagz_QPI}
\end{center}
\end{figure}

Different from nonmagnetic impurities, a weak magnetic impurity has
very little effect on the charge density of the system, namely,
instead of having
$\delta\rho_\uparrow(\mathbf{q},\omega)=\delta\rho_\downarrow(\mathbf{q},\omega)$
as in the nonmagnetic impurity case,  we  have
$\delta\rho_\uparrow(\mathbf{q},\omega)\approx-\delta\rho_\downarrow(\mathbf{q},\omega)$
. This effect can be easily understood. Suppose we are considering
an impurity moment along the $z$-direction, then the spin-up
electrons and spin-down electrons see two scattering potentials of
opposite signs. In the lowest order of perturbation theory, the
scattering amplitude of the spin-up and spin-down electrons thus
differ by a minus sign so that the total interference pattern of the
charge density vanishes almost everywhere. The same argument no
longer holds if higher orders of perturbation are included. For the
model considered here, we can explicitly prove the above statement.
Assuming $V\ll\omega$, the approximation $T(\omega)\approx\hat{V}$
becomes sufficiently accurate. In this case (impurity moment along
$z$-direction), we have\begin{widetext}\bea
\nonumber&&{\text{Tr}}[\delta{G}(\mathbf{q},\omega)]\approx\int\frac{d^2k}{(2\pi)^2}
\text{Tr}[G_0(\mathbf{k},\omega)\hat{V}G_0(\mathbf{k+q},\omega)]\\
\nonumber&=&V\int\frac{d^2k}{(2\pi)^2}\frac{\text{Tr}
[(\omega\sigma_0-k_y\sigma_x+k_x\sigma_y+\frac{\lambda}{2}
(k_+^3+k_-^3)\sigma_z)\sigma_z(\omega\sigma_0-(k_y+q_y)\sigma_x+(k_x+q_x)\sigma_y+\frac{\lambda}{2}((k+q)_+^3+(k+q)_-^3)\sigma_z)]}{((\omega+i\eta)^2-\epsilon_+^2(\mathbf{k}))((\omega+i\eta)^2-\epsilon_+^2(\mathbf{k+q}))}\\
\nonumber&=&V\int\frac{d^2k}{(2\pi)^2}(2\frac{k_+^3+k_-^3+(k+q)_+^3+(k+q)_-^3+ik_y(k_x+q_x)-ik_x(k_y+q_y)}{((\omega+i\eta)^2-\epsilon_+^2(\mathbf{k}))((\omega+i\eta)^2-\epsilon_+^2(\mathbf{k+q}))})\\
&=&0.\eea
\end{widetext}
%From the second last line to the last line:
The last equality is achieved by   shifting the origin to
$(q_x,q_y)$, changing the integrated variables $\mathbf{k}$ to
$-\mathbf{k}$, and taking the advantage that
$\epsilon_+(\mathbf{k})=\epsilon_+(\mathbf{-k})$. Similar
derivations hold for the impurity moment along $x$ and
$y$-directions. If the second order term $\Cal{O}(V^2)$ is included
in the $T$-matrix, the cancellation becomes no longer valid, and
there is indeed small but finite charge LDOS pattern in the system.
 In Fig.~\ref{fig:tmagz_QPI}, we plot the numerical results of
$\delta\rho(\bq,\omega)$ at $\omega=0.05,0.3$. It is clear that The
amplitude of charge density variation by magnetic impurities in
Fig.~\ref{fig:tmagz_QPI} is two orders of magnitude smaller than that
shown in Fig.~\ref{fig:nonmag_QPI} by nonmagnetic impurities.

Therefore, for the magnetic impurity case, we should choose a
time-reversal breaking observable to study the interference, and a
natural choice is the spin local density of states (SLDOS),
defined
by\bea\vec{S}(\br,\omega)=-\frac{1}{\pi}\text{Im}[\int{dt}\theta(t)
\langle{}c_\alpha(\br,t)\vec{\sigma}^{\alpha\beta}c^\dag_\beta(\br,0)
\rangle{}e^{i\omega{t}}],\eea where $c^\dagger_{\alpha}(\br,t)$
creates an electron with spin polarization $\alpha$ at position
$\br$ and time $t$. From now on we will only focus on the FT of
the $z$-component SLDOS.

\begin{figure}[tbh]
\begin{center}
\includegraphics[scale=0.45]{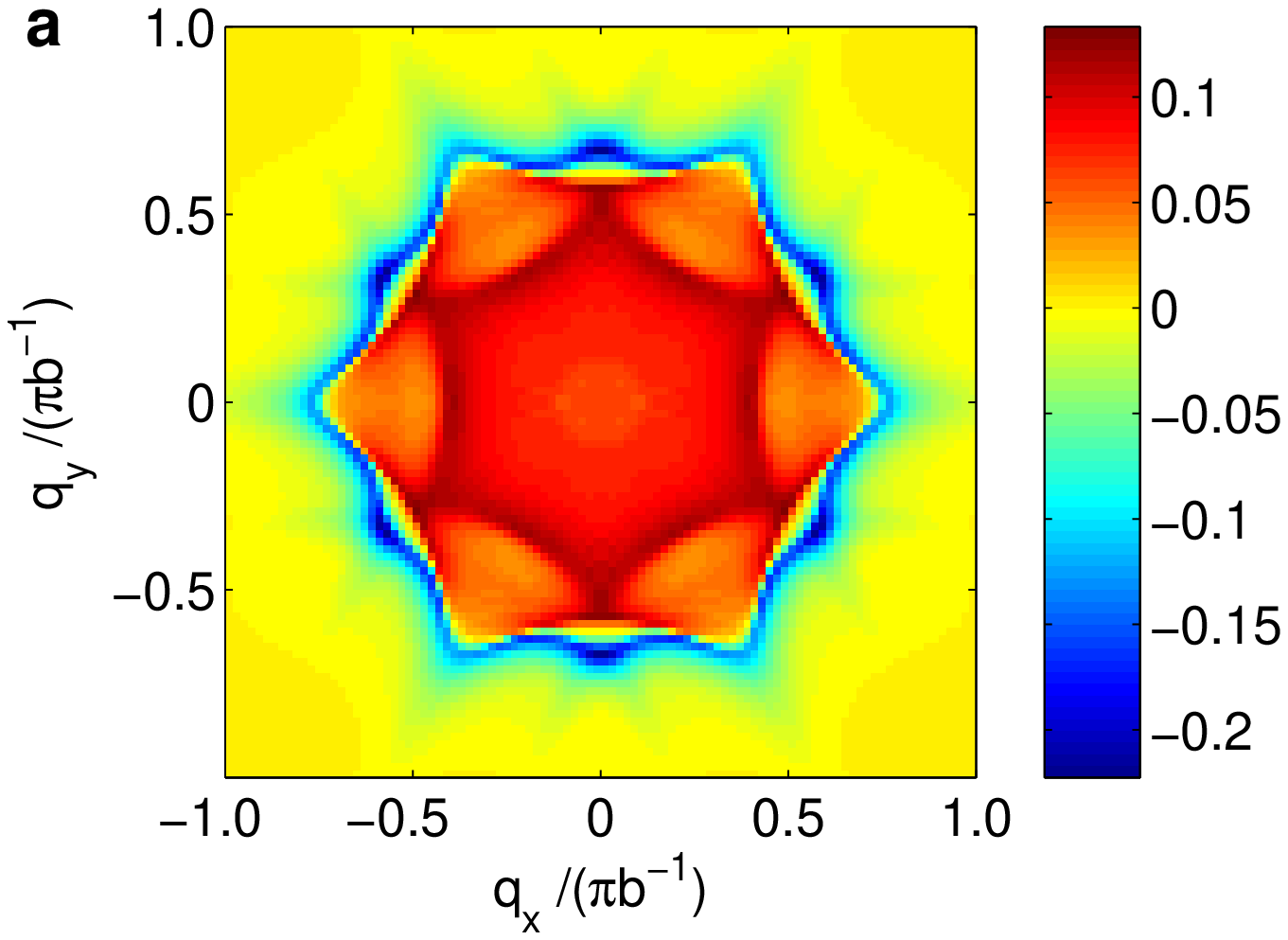}
\includegraphics[scale=0.45]{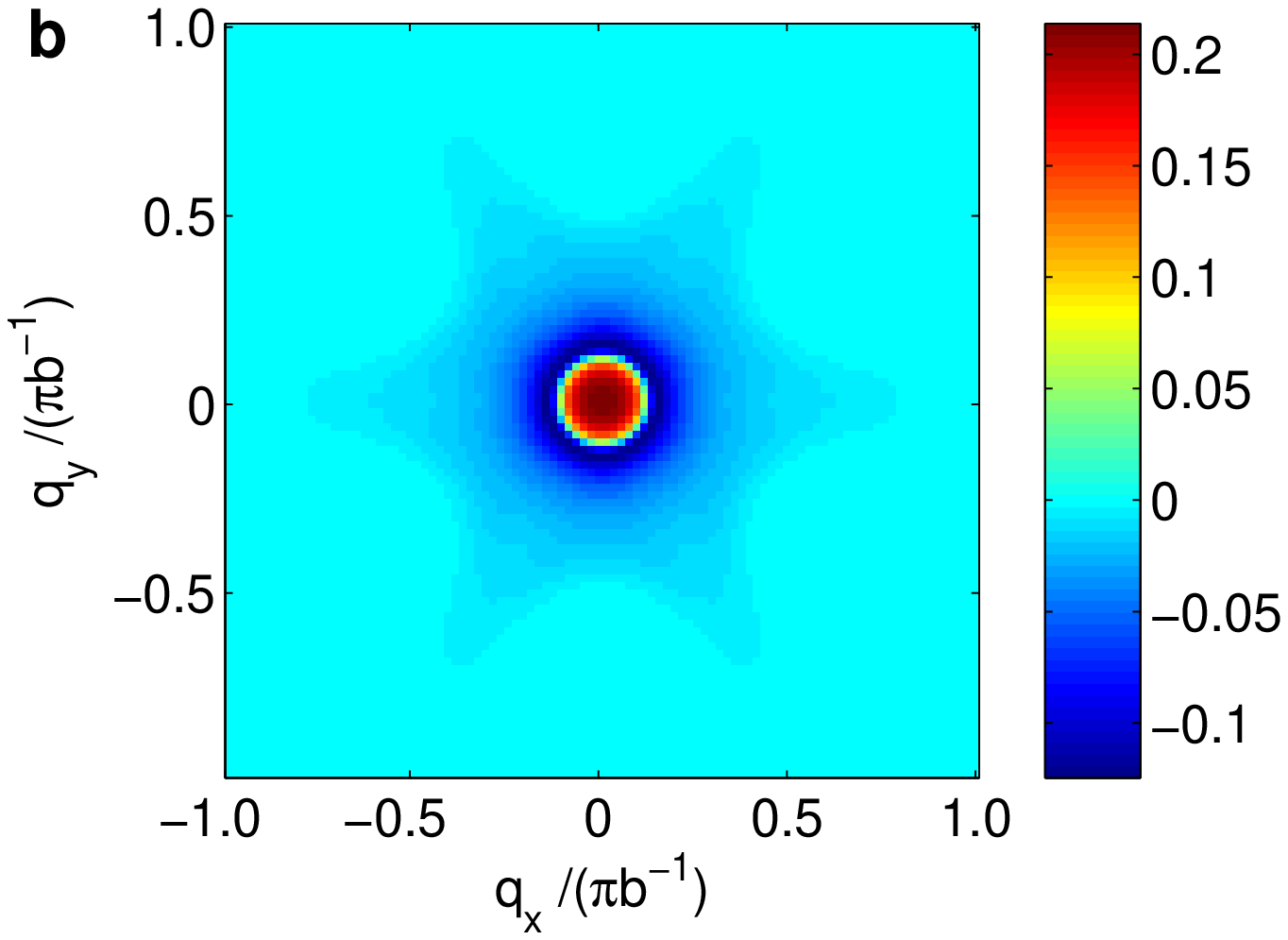}
\caption{The real part of the Fourier transform of spin local
density of states in the case of single magnetic point impurity
with its spin polarized along the z-axis at (a) $\omega=0.3eV$,
and (b) $\omega=0.05eV$.} \label{fig:dmagz_QPI}
\end{center}
\end{figure}

In the case of nonmagnetic impurity, we have demonstrated the
absence of interference between $|\mathbf{k},\uparrow\rangle$ and
$|\mathbf{-k},\downarrow\rangle$, which form a time-reversal pair.
Physically, a time-reversal breaker such as a magnetic impurity can
lift this ban on the backscattering. Similar to
Eq.~(\ref{eq:forbidden}), it is easy to show that
$\langle{-\mathbf{k}},\downarrow|\hat{V}|\mathbf{k},\uparrow\rangle\neq
0$ due to $\Cal{T}\sigma_i\Cal{T}^{-1}=-\sigma_i$. This feature is
universal in all of our figures for magnetic impurity. Taking
Fig.~\ref{fig:dmagz_QPI}(a) as an example, we can compare it with
Fig.~\ref{fig:nonmag_QPI}(a) and notice that although they have
common features, the points in the FT-SLDOS that associate with the
$2\bk_F$ backscattering scattering vectors is only present
($\pm\bq_{14},\pm\bq_{25},\pm\bq_{36}$) (see Fig.~\ref{fig:CCE_EW})
%$\mathbf{q_{14},q_{25},q_{36}}$
 in the magnetic scattering. We can also compare
   Fig.~\ref{fig:dmagz_QPI}(b) for magnetic scattering with
Fig.~\ref{fig:nonmag_QPI}(b) for nonmagnetic scattering when
$\omega=0.05$eV. In the latter case, the interference strength
universally decays quickly after reaching the boundary of the
circle; while in the former case, the interference strength reaches
a negative peak   across the boundary, indicating a scattering that
flips spin moments of the quasiparticles.

\begin{figure}[tbh]
\begin{center}
\includegraphics[scale=0.4]{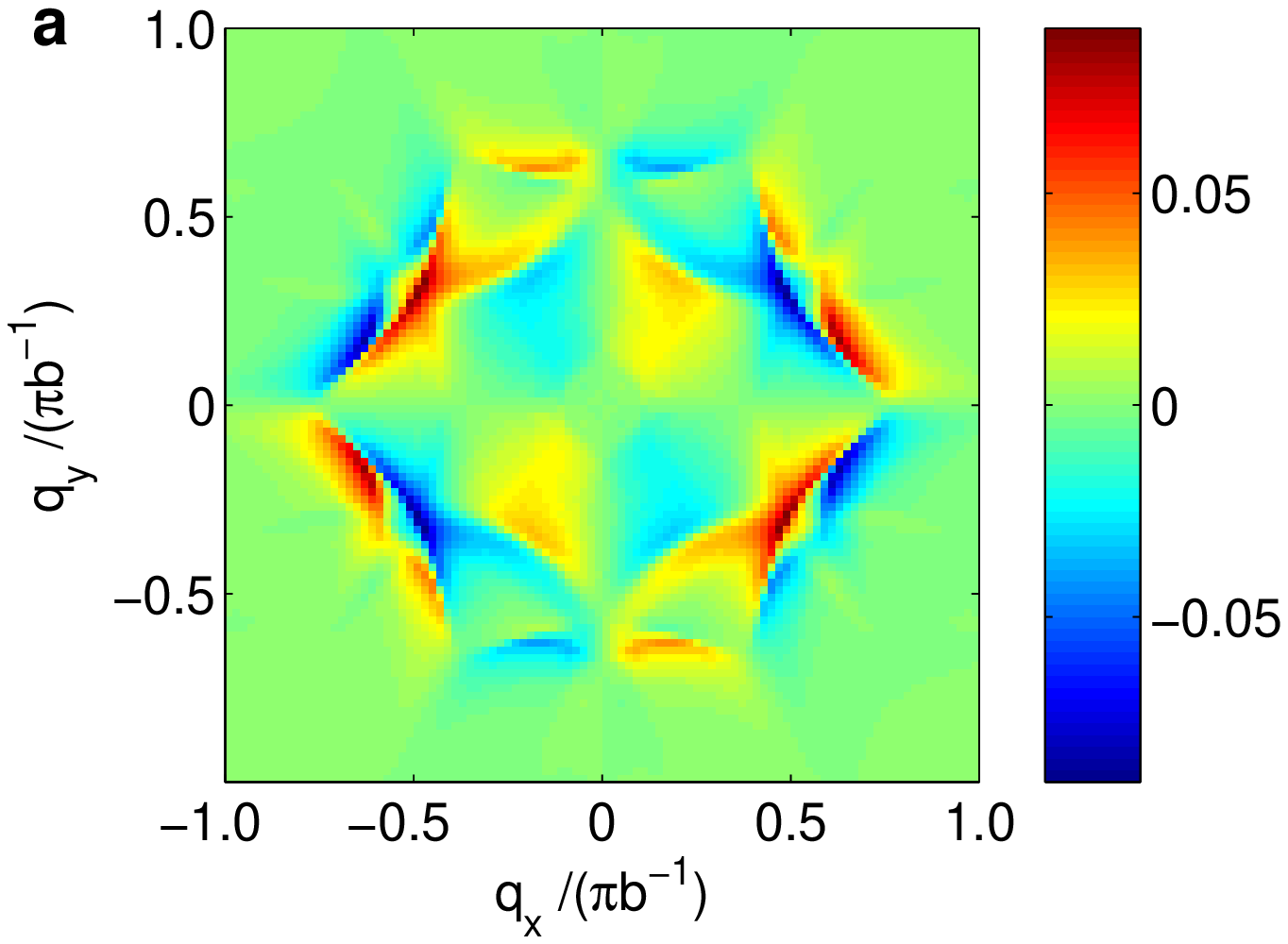}
\includegraphics[scale=0.4]{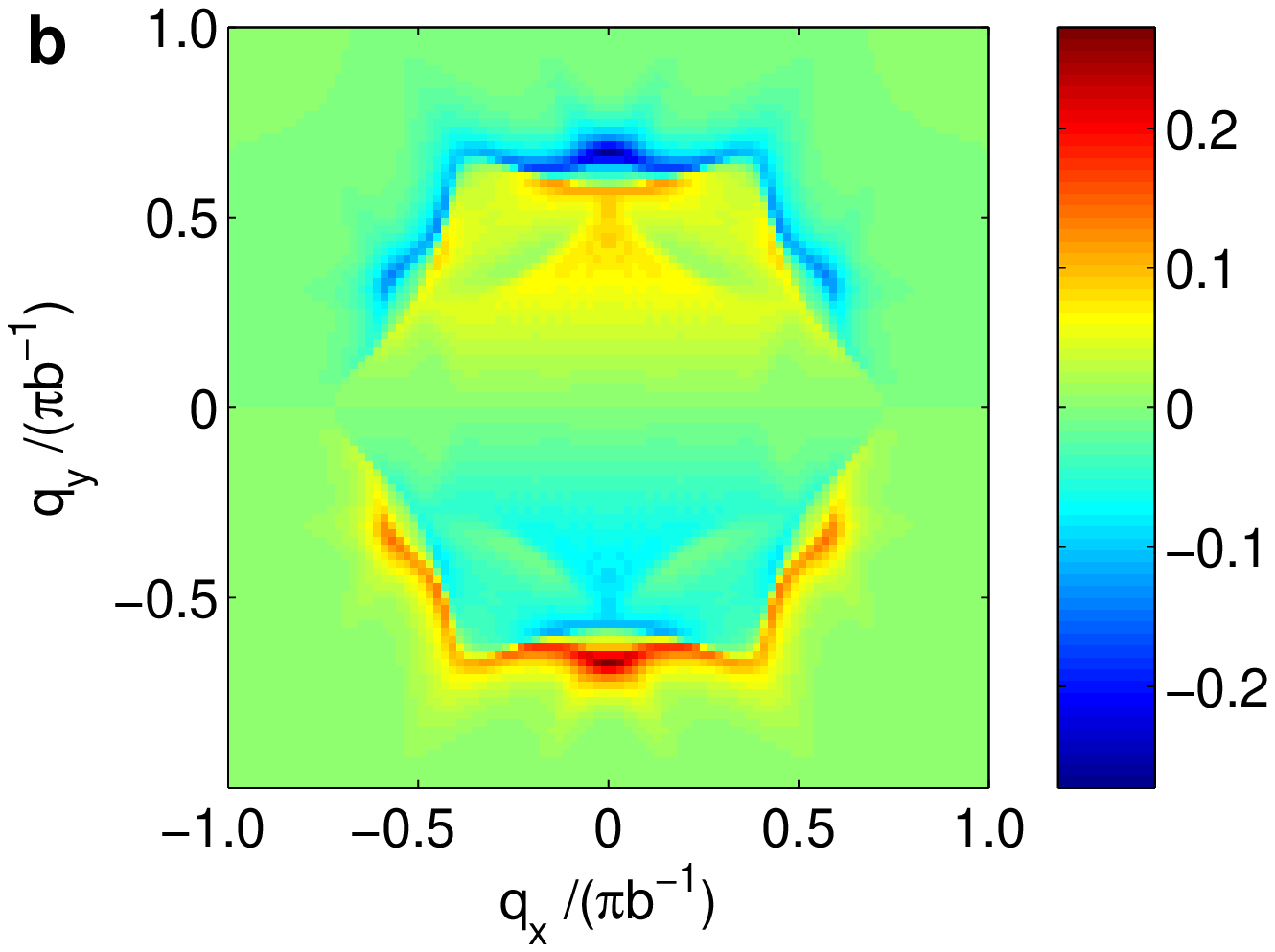}
\caption{The (a) real part and the (b) imaginary part of the
Fourier transform of spin local density of states in the case of
single magnetic point impurity with its spin polarized along the
$y$-axis at $\omega=0.3eV$.} \label{fig:dmagy_QPI}
\end{center}
\end{figure}

\begin{figure}[tbh]
\begin{center}
\includegraphics[scale=0.4]{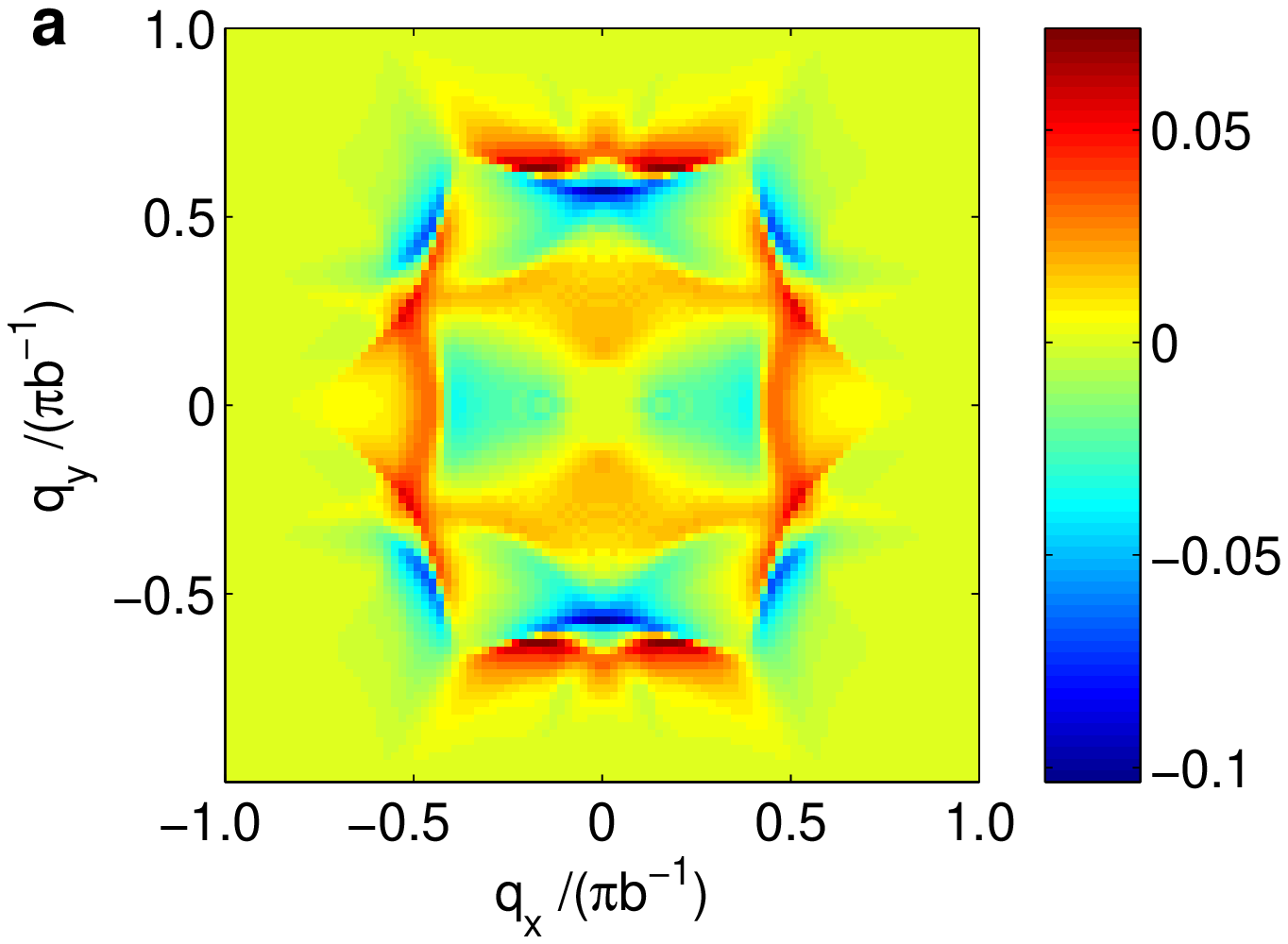}
\includegraphics[scale=0.4]{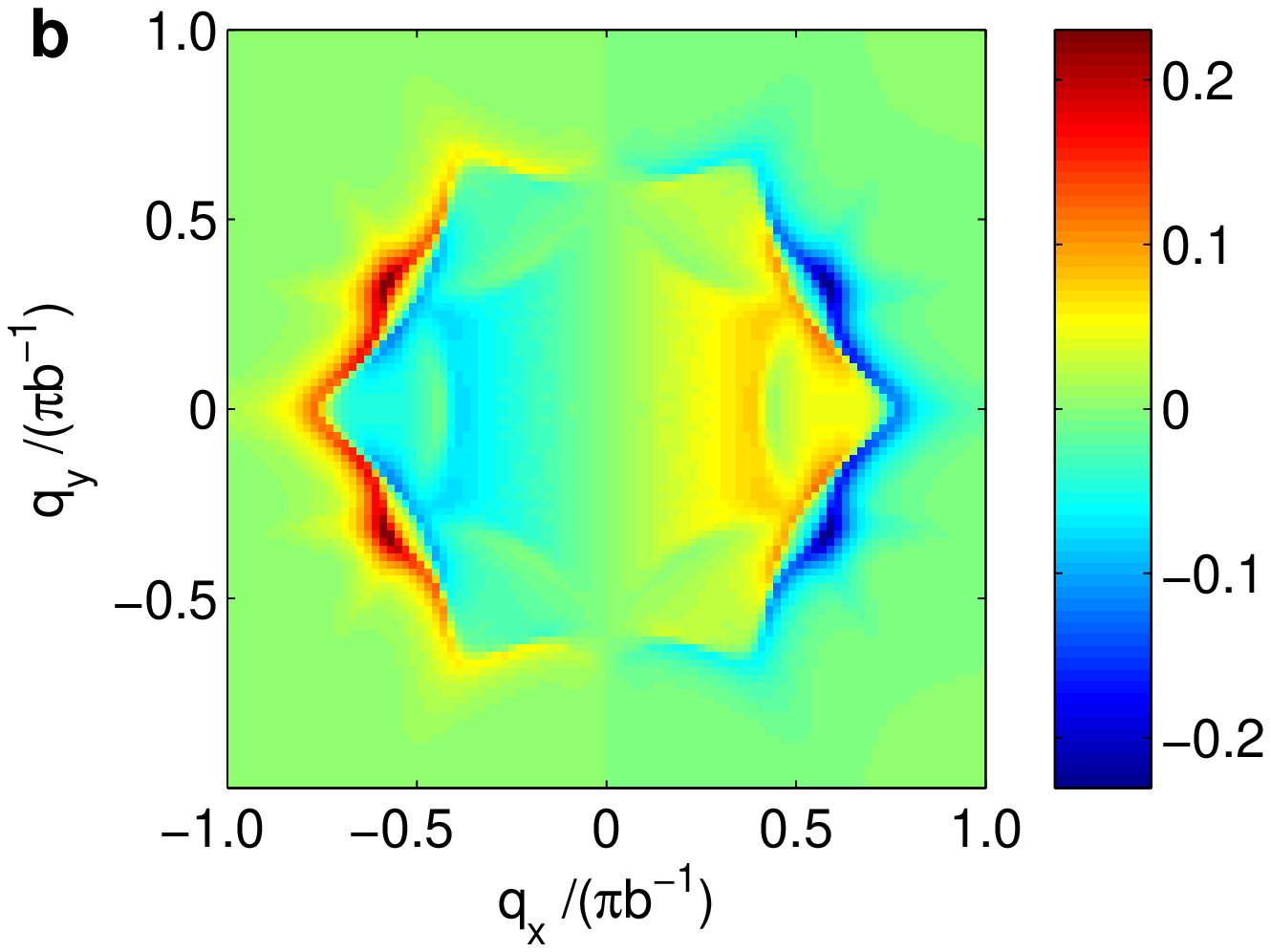}
\caption{The (a) real part and the (b) imaginary part of the
Fourier transform of spin local density of states in the case of
single magnetic point impurity with its spin polarized along the
$x$-axis at $\omega=0.3eV$.} \label{fig:dmagx_QPI}
\end{center}
\end{figure}

Now, we discuss the QPI by magnetic impurities with in-plane
magnetic moments. In this case, a unique feature rises in the
FT-SLDOS. As shown in Fig.~\ref{fig:dmagy_QPI}and in
Fig.~\ref{fig:dmagx_QPI}, at $\omega=0.3$ we plot two figures,
which correspond to the real and imaginary parts of the FT-SLDOS
separately.  Similar to the LDOS, the real and imaginary parts
correspond to the symmetric and antisymmetric parts of
$S_z(x,y,t)$ respectively. For magnetic impurity with magnetic
moment along z axis, the symmetric part dominates and the
antisymmetric part is either vanishing or orders of magnitude
smaller than the symmetric part. However, here as shown in
Fig.~\ref{fig:dmagy_QPI}, the antisymmetric part is about three
times larger than the symmetric part. the result can be understood
as follows. An inversion transformation in a two dimensional
plane, i.e. $(x,y)\rightarrow(-x,-y)$ takes
$\hat{\sigma}_z(x,y,t)\rightarrow \hat\sigma_z(-x,-y,t)$ and
$\hat\sigma_{x,y}(-x,-y,t)\rightarrow-\hat \sigma_{x,y}(-x,-y,t)$.
Therefore, under this transformation, the Hamiltonian without the
warping term in the presence of magnetic impurities with in-plane
magnetic moments transforms as $H(V_0)\rightarrow H(-V_0)$, where
$V_0$ is the coupling strength of magnetic impurity. Thus, from
this symmetry, if
 we consider $S_z(x,y,t)$ as function of $V_0$ as well, we have
 $S_z(x,y,t,V_0) = S_z(-x,-y,t,-V_0)$. Therefore, the first order correction
 from the scattering potential vanishes for the symmetry part. In the presence of the warping term,
 there is no such an exact symmetry argument. Nevertheless, the symmetric part is still much smaller than the antisymmetric part. In
 the following, we will first focus on the antisymmetric part.

Fig.~\ref{fig:dmagy_QPI}(b) shows the (antisymmetric)
FT-interference pattern for the impurity moment along the $y$-axis
at $\omega=0.3$eV. We find that the strongest interference appears
at wave vector $\pm\mathbf{q}_{51}$ in Fig.~\ref{fig:dmagy_QPI}(b)
($q_{ij}$ is defined in Fig.~\ref{fig:CCE_EW}). Moreover,
$\mathbf{q}_{13}$ and $\mathbf{q}_{35}$ do not present as strong
peaks,  in contrast with the cases of the nonmagnetic impurity   and
the magnetic impurity spin along $z$-axis. In addition, a remarkable
feature in the interference pattern is that
$S_z^A(\mathbf{q},\omega)$ is
 zero on the line $q_y=0$.
This is caused by an exact symmetry of the %symmetry of the
system which dictates $S_z(x,y,t)=-S_z(x,-y,t)$. This point will be
discussed later in length. Fig.~\ref{fig:dmagx_QPI}(b) shows the
(antisymmetric) FT-interference pattern for the impurity spin along
the $x$-axis at $\omega=0.3$eV. We can see that the strongest
interference is associated with the vertex-to-vertex wave vectors
$\mathbf{q}_{13}$ and $\mathbf{q}_{35}$. The strong peak at
$\mathbf{q}_{51}$ does not appear and we have $S^A_z(0,q_y,\omega)$
vanishing. This result stems from an approximate equality
$S_z(x,y,t)\approx{}S_z(x,-y,t)$, a point of which will be discussed
next.

We can understand above detailed features in the SLDOS from the
symmetry analysis of the model. The model  obviously has the
time-reversal symmetry and the three-fold rotation symmetry.
Moreover, the model also preserves the $y\rightarrow-y$ mirror
symmetry ($m_y$) but breaks the $x\rightarrow-x$ mirror symmetry
($m_x$), as can be seen in the warping term. Explicitly, the $m_x$
operator takes $k_\pm$ to $k_\mp$ and $\sigma_z$ to $-\sigma_z$,
which changes the sign of the warping term. Now, let us consider
the system in the presence of a magnetic impurity with its spin
along $y$-axis. Since $s_y\rightarrow s_y$ under ${m_y}$, the
whole system still preserves the mirror symmetry $m_y$. This
symmetry directly leads to \be S_z(x,y,\omega)=-S_z(x,-y,\omega).
\ee  This symmetry property is clearly demonstrated in
Figs.~\ref{fig:dmagy_QPI}(a) and (b). On the other hand, if the
impurity spin is fixed along the $x$-direction, the system does
NOT have $m_x$ symmetry and we have
$S_z(x,y,\omega)\neq-S_z(-x,y,\omega)$.  This feature is also
demonstrated in Fig.~\ref{fig:dmagx_QPI}(a). If we had
$S_z(x,y,\omega)=-S_z(-x,y,\omega)$, we should have
$S_z^{A(S)}(q_x,q_y,\omega)=-S_z^{A(S)}(-q_x,q_y,\omega)$ or
$S_z(x,y,\omega=S_z(-x,y,\omega)=0$. However, in
Fig.~\ref{fig:dmagx_QPI}(a), it is clear that
$S_z^{S}(q_x,q_y,\omega)=S_z^{S}(-q_x,q_y,\omega)\neq0$.

The above symmetry is a very important property of the model. In
fact, to simply account for the shape of FS, we may also
artificially make the Fermi velocity strongly angle dependent while
keeping the same spin texture where all spins on the FS are in-plane
without tilting. For instance, we can write \be
\tilde{H}(\bk)=v(\bk)(k_x\sigma_y-k_y\sigma_x)+\frac{k^2}{2m^*},
\label{model2} \ee where
$v(\bk)=\sqrt{v^2+\lambda^2k^4\sin^2(3\theta)}$, with $\theta$ being
the azimuthal angle with respect to $x$ axis ($\Gamma M$). This
model (the in-plane model) has the same dispersion as the model in
Eq.\ref{eq:model}, but has only in-plane spin texture. The
symmetries of the SLDOS here can help us distinguish these two
models. For example, one can check these two equations
experimentally: $S_z(x,y,\omega)=-S_z(x,-y,\omega)$ for impurity
spin polarized along $y$ axis and
$S_z(x,y,\omega)=-S_z(-x,y,\omega)$ for impurity spin polarized
along $x$ axis. If both are held, then the in-plane model suffices;
but if only one is held, we may need an out of plane spin (warping)
term. In Table.\ref{table1}, we list the property of SLDOS in the
two models, Eq.(\ref{eq:model}) and Eq.(\ref{model2}), in the presence
of different types of impurities and under basic symmetry
operations.

\begin{table}
\begin{tabular}{cc}
 \begin{tabular}{|c|c|c|c|}
  \hline
  $S_z$ & $m_x$ & $m_y$ & $C_3$ \\
  \hline
  $s_x$ & $\times$ & $\approx1$ & $\times$ \\
  $s_y$ & $\times$ & -1 & $\times$ \\
  $s_z$ & $\approx1$ & $\approx1$ & 1 \\
  \hline
\end{tabular}
&
 \begin{tabular}{|c|c|c|c|}
  \hline
  $S_z$ & $m_x$ & $m_y$ & $C_3$ \\
  \hline
  $s_x$ & -1 & $\approx1$ & $\times$ \\
  $s_y$ & $\approx1$ & -1 & $\times$ \\
  $s_z$ & $\approx1$ & $\approx1$ & 1 \\
  \hline
\end{tabular}\\
(a)& (b)\\
\end{tabular}
\caption{The symmetry of $S_z(x,y,t)$ under symmetry operations of
mirror-x ($m_x$), mirror-y ($m_y$) and three-fold rotation about
z-axis ($C_3$) with impurity spin along three axes. (a) is for the
model in Eq.(\ref{eq:model}) and (b) the model in
Eq.(\ref{model2}). '$1$' means symmetric; '$-1$' means
antisymmetric and '$\times$' means neither of the above. The
'$\approx$' means it is symmetric (antisymmetric) in the weak
impurity strength approximation.} \label{table1}\end{table}

\subsection{Nonmagnetic edge impurity}
Step atomic roughness on a surface may be locally idealized into
an edge impurity, that is, an infinite line with different but
uniform potential on two sides. An edge impurity in a 2D
conventional Fermi gas is known to give rise to Friedel
oscillation {\it at fixed energy} in the LDOS. This oscillation can
simply be understood as an interference pattern between the
incoming plane wave and the reflected wave by the 1D edge. The
major contribution comes from the two opposite $\mathbf{k}$-points
on the constant energy contour, $\pm\bk_F$, and the oscillation
has the wavenumber $2|\bk_F|$ while decaying as a form
$1/\sqrt{d}$ where $d$ is the distance from the edge
impurity\cite{crommie93}. The same picture is no longer valid if
the state at $\bk$ and $-\bk$ do not scatter with each other, a
case for the surface states of a 3D topological insulator where
the backscattering is forbidden by the time-reversal symmetry.
Therefore the oscillation is expected to decay much faster and
thus practically absent in an STM experiment. The 'absence' of the
Friedel oscillation is considered as a sign of (spin) helical
Dirac Fermion systems. However, the
 oscillation has been observed in STM experiments\cite{alpichshev09}. The apparent discrepancy between
theory and experiment was soon claimed to be superficial and
explained by the hexagram-shape of the FS\cite{fu09}. In this
subsection an exact calculation is performed to test this physical
picture.

We consider that the edge impurity is fixed along $y$ axis and the
system has zero potential for $x<0$ and uniform potential $V$ for
$x>0$. A general quantum state on the left hand side (LHS)  takes
the form \be
\psi(k_x,k_y;x,y)=\frac{\phi_0(k_x,k_y;x,y)+r\phi_0(-k_x,k_y;x,y)}
{\sqrt{1+|r|^2}}, \ee and the LDOS is \be\label{eq:edge_LDOS}
\rho(x,\omega)=\int_{k_x>0}\frac{d^2k}{(2\pi)^2}|\psi(k_x,k_y;x,y)|^2
\delta(\omega-\epsilon_+(k_x,k_y)). \ee The reflection amplitude
$r$ can be obtained together with the transmission amplitude $t$
by matching the boundary condition at the edge, namely, \be
\phi_0(k_x,k_y;0,y)+r\phi_0(-k_x,k_y;0,y)=t\phi_0(k''_x,k_y;0,y),
\ee where $k''_x$ is  fixed by the energy conservation
$\epsilon(k_x,k_y)=\epsilon(k''_x,k_y)-V$.

\begin{figure}[tbh]
\begin{center}
\includegraphics[scale=0.45]{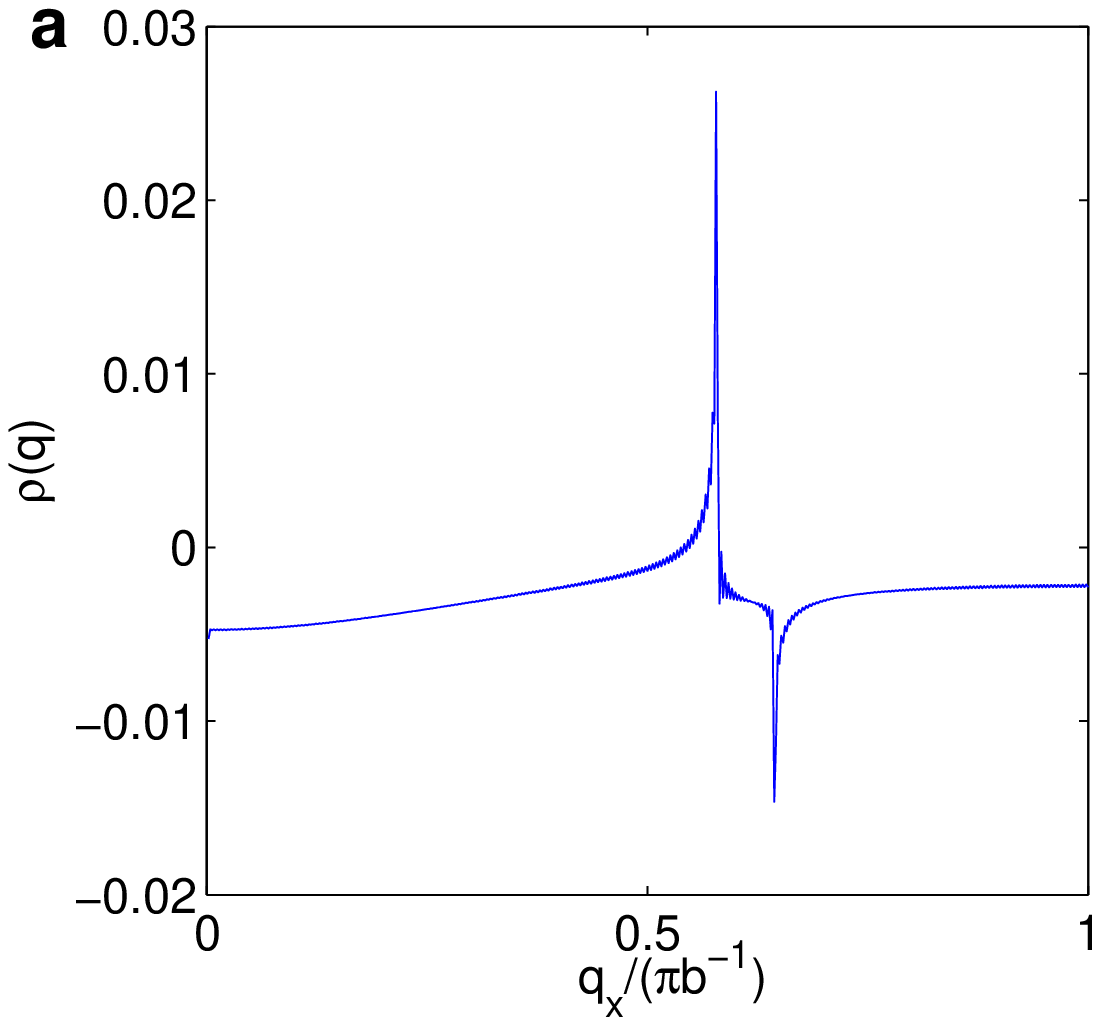}
\includegraphics[scale=0.45]{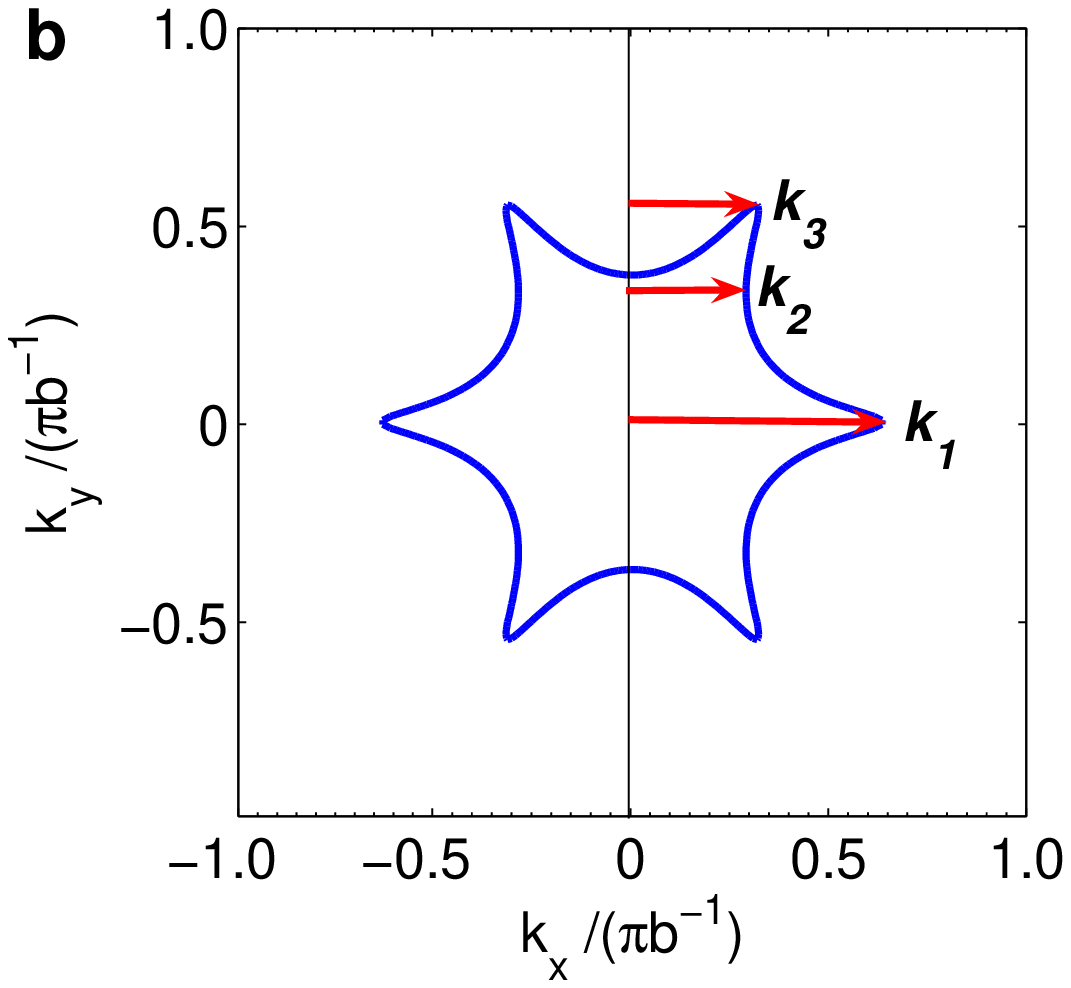}
\caption{(a)The Fourier transform of the edge impurity ($V=-0.1$)
interference pattern. (b)The three $\bk$'s that dominate the
interference pattern on the energy contour at $\omega=0.5$eV.}
\label{fig:FT_edge_omega05}
\end{center}
\end{figure}

\begin{figure}[tbh]
\begin{center}
\includegraphics[scale=0.45]{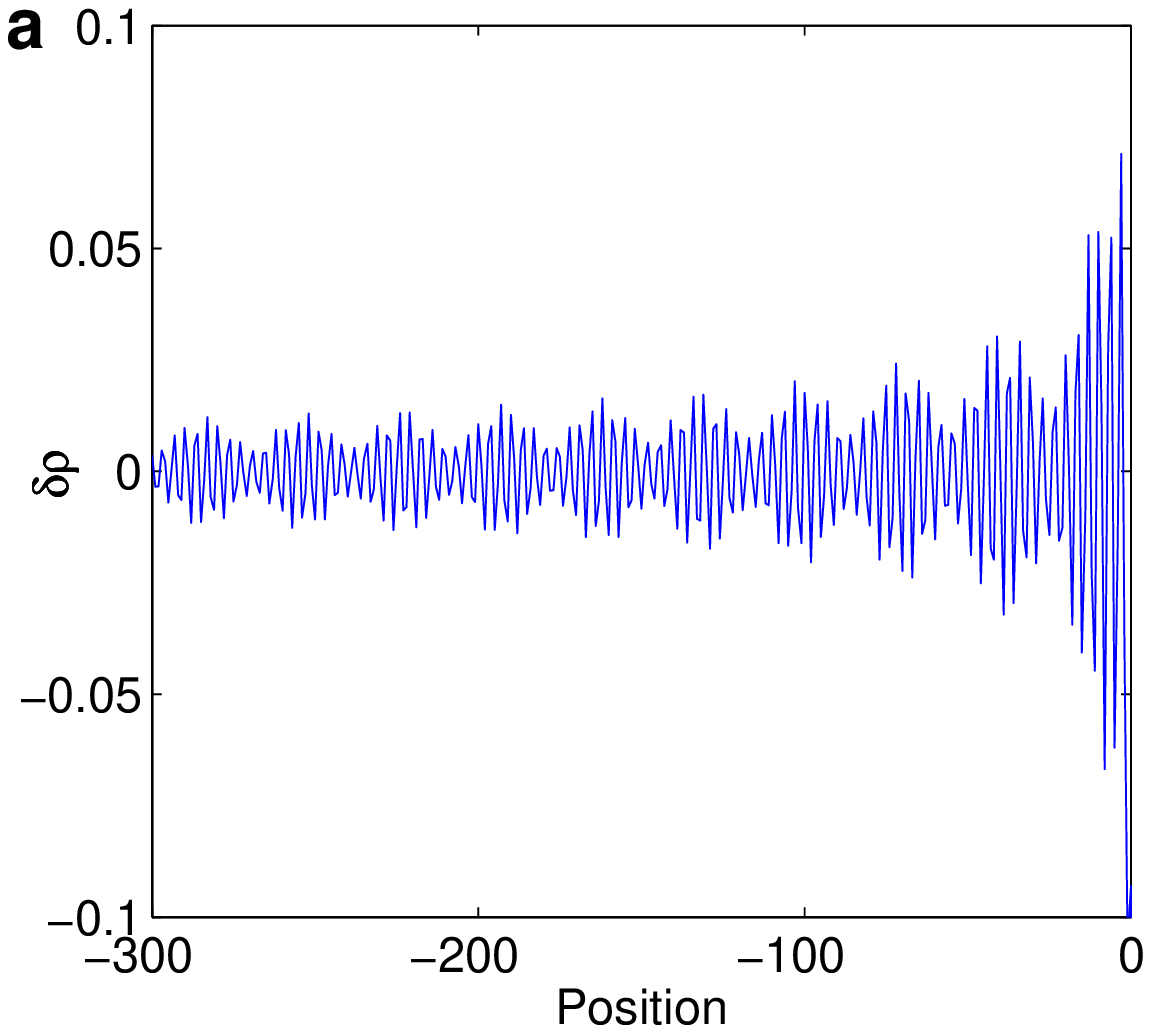}
\includegraphics[scale=0.45]{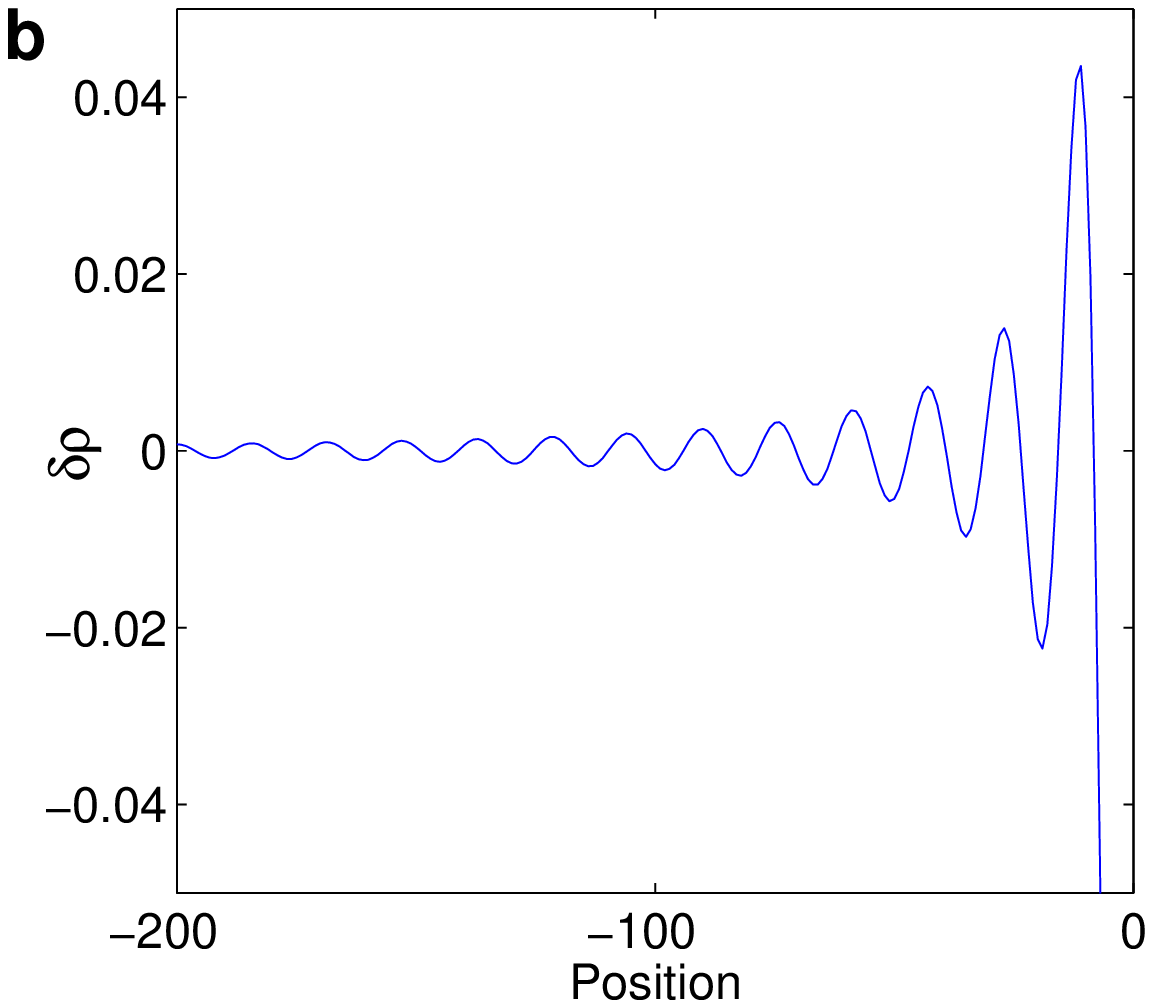}
\caption{The real space interference pattern for the edge impurity
($V=-0.1$) at (a) $\omega=0.5$eV and (b) $\omega=0.05$eV. The
density fluctuation $\delta\rho$ is defined as
$\delta\rho=\rho-\rho_0=\rho-1$. The position $x$ is in units of $b$.} \label{fig:real_edge}
\end{center}
\end{figure}

Fig.~\ref{fig:FT_edge_omega05}(a) shows the FT-LDOS for the LHS of
the edge  impurity at $\omega=0.5$. We can clearly identify the
two peaks in the interference associated with $q_x=2\bk_{2}$ and
$q_x=2\bk_3$, defined in Fig.~\ref{fig:FT_edge_omega05}(b). No
feature is present at $q_x=2\bk_1$, reflecting the absence of
backscattering. The spatial dependence of the oscillation, a real
space LDOS,  is given in Fig.~\ref{fig:real_edge}(a). A clear
beating pattern can be seen with spatial period $\sim
(\bk_3-\bk_2)^{-1}$. The oscillation decays like $1/|x|^{\alpha}$
where $\alpha\sim0.46$, qualitatively matching the theoretical
prediction in the large $|x|$ limit\cite{crommie93,fu09b}. When
$|x|$ is large enough, the stationary points approximation tells
us that, if the edge impurity is along the $y$-axis, the
interference pattern is dominated by the $\bk$-points where $k_x$
reaches local minimum or maximum. In our model, $\bk_{2(3)}$ are
the points corresponding to the minimum (maximum) of $k_x$ on the
contour of constant energy. However, The existence of such extrema
depends on $\omega$. If $\omega$ is small enough, the extrema
$\bk_{2,3}$ disappear and we are left with only $\bk_1$. Since
$\bk_1$ is not allowed to scatter with its time-reversal partner,
the decaying of Friedel oscillation
 becomes $|x|^{-3/2}$ at large $|x|$
[see Fig.~\ref{fig:real_edge}(b)]. Therefore, there  is no
universal function for the oscillation decay.  The decay depends
on the values of parameters. There are two inherent length scales
in the model: $b=\sqrt{\lambda/v}$ and $b'=v/\omega$. If
$b>1.48b'$, the energy contour is a hexagram and an $1/\sqrt{|x|}$
decay of the oscillation appears, while if $b\ll{b}'$, we have a
nearly circular FS and the decay of oscillation takes the form
$\rho(x)\sim|x|^{-3/2}$. In the intermediate range, the
oscillation varies. For example, at $b=1.2b'$ ($\omega=0.3$), the oscillation
decays exponentially for $|x|<100b$ but close to  $|x|^{-3/2}$ for
$|x|>200b$.
\begin{figure}[tbh]
\begin{center}
\includegraphics[scale=0.3]{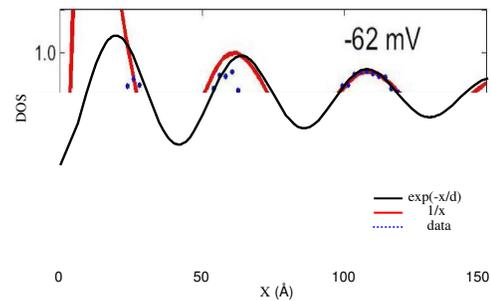}
\caption{Fitting the experimental data of
Ref.\onlinecite{alpichshev09} using different oscillating
functions. The experimental energy -62meV corresponds to
$\sim0.25$eV in our units. In the exponential fit, $d=107\r{A}$.}
\label{fig:exp_exp}
\end{center}
\end{figure}

\section{Discussion and conclusion}
The model we have solved produces interference patterns that have
enough  features to be associated with the topology-protected
surface states and the effects of the hexagonal warping term in 3D
topological insulators. However, in order to be more careful before
making conclusions, there are two more remarks we would like to
mention here.

%[Coulomb interaction, order]
(i) In our calculations, we neglected the possibility of any
ordering due to interaction-induced FS instability. This is valid as
long as there is no significant FS nesting vector\cite{fu09}. In
addition, we do not expect strong electron-electron interaction
based on the following observation. In experiments on topological
insulators, the Fermi level of the sample in general is closer to
the bottom of the conduction band and is far away from the Dirac
point. Such a system with finite density of states may provide
enough screening effect to Coulomb interaction between surface
electrons. Moreover, attempting to tune the Fermi level lower by a
metallic gate may also lead to the same phenomenon, turning
interaction between electrons into irrelevant regime.

(ii)In real systems, there is no 'purely magnetic' impurity. A
magnetic impurity should also have a non-magnetic component. This
fact does not change our results obtained for magnetic impurities.
In the parameter region we choose, the weak impurity approximation
is always valid (see a detailed discussion of this approximation in
the Appendix), the non-magnetic impurity only leads to the charge
density modulation and  has little  effect on the SLDOS. Namely,
  the magnetic part of impurity is solely responsible for the SLDOS.

%[Remarks in experiments]
(iii) As we noticed in section III A, the STM experiment done  by
Zhang et al.\cite{zhang09} on [111] surface of Bi$_2$Te$_3$
exhibited six peaks in FT-LDOS for the case of nonmagnetic
impurities. The experimental result differs from our results shown
in Fig.~\ref{fig:nonmag_QPI}(a) by a 30 degrees of rotation.
However, this discrepancy can be understood by noticing that in the
energy range where they observed the clear interference patterns
(50meV$\sim$400meV), the surface density of states are mixed with
bulk states along $\Gamma M$. Consequently, due to the superposition
of waves with various wavelengths the interference patterns are
simply smeared out in these regions. Instead of a full FS we
considered here, the dominant interference patterns are then from
other unmixed parts of the FS, \ie, the parts along $\Gamma K$.

(iv)  In an STM experiment done by Alpichshev et
al.\cite{alpichshev09}, the decaying behavior of the Friedel
oscillation was claimed to be $1/|x|$. However,  in the case of 1D
edge impurities, our calculation shows $1/|x|^{1/2}$ behavior if the
FS shape is dominated by the warping term, and $|x|^{-3/2}$ if the
warping term is negligible.  We believe there are two possible
sources of the discrepancy. First,  we notice that a simple fitting
to the first several periods of oscillation is not enough to
determine the decaying behavior. In Fig.\ref{fig:exp_exp}, we show
that the data in Ref.\onlinecite{alpichshev09} can also be well
fitted using an exponentially decaying function, as opposed to the
$1/|x|$-type fit used in Ref.\onlinecite{alpichshev09}. Second, the
experimental measurements are not a pure surface effect. There are
bulk electrons in the nearby conduction band which can cause
different decaying behavior and complicate the issue. More future
experimental measurements are necessary to resolve the issue and
test the theoretical predictions.

(v) We also notice that a similar theoretical work\cite{lee09}
focusing solely on nonmagnetic impurity was posted online recently,
which suggests the six peaks at fixed $\bq$'s that correspond to the
scattering vectors connecting between second neighbor of the convex
parts of the FS dominate in the QPI patterns. Their results are
consistent with our calculations since their results, according to
the energy unit in our paper, are obtained at $\omega=0.375eV$.
However, our results suggest that the relative strength between the
interference at $\bq$'s connecting next nearest neighboring vertices
(e.g., $\bq_{35}$ and the interference at $\bq$'s connecting next
nearest neighboring arc-centers (e.g., $\bq_{2'4'}$ in the QPI
patterns is quite subtle and depends on energy. Therefore, a full
$T$ matrix calculation is necessary in calculating the QPI patterns.

In conclusion, we have  investigated the quasiparticle scattering in
a 2D helical liquid in the presence of nonmagnetic/magnetic point
impurity or an nonmagnetic edge impurity. The inclusion of the
hexagonal warping term in our system not only inherits the nature of
the $k$-linear helical liquid but also sharpens our features
mentioned above by distorting the shape of the FS. More importantly,
it requires an out of plane spin texture and can be distinguished
from other systems with examination of the mirror symmetries when
the magnetic point impurity with in-plane spin moment is present.
The absence (presence) of spots in FT-LDOS (FT-SLDOS), corresponding
to the backscattering interference, are the essential features to
confirm the topological nature of the helical liquid. The results in
our work, as may be detected by STM experiments, can be a useful
quantum signature, which is uniquely associated with this new phase
of matter, a 3D topological insulator.

%[*matrix element effect, FS re-organized by certain ordering]

\begin{acknowledgments}
The authors thank Liang~Fu for his insights and stimulating
discussion and H.~Yao for useful conversation.
\end{acknowledgments}
%\vspace{-0.4cm}

\appendix
\section{The weak impurity strength approximation}

For the parameter we used throughout the paper, the scattering
strength is relatively small (i.e., $V_0\rho(\omega)\ll1$). In this
limit the approximation $T(\omega)\approx\hat{V}$ is considerably
accurate (less than 3\% error in our case), and many approximate
equalities may be derived thereof. This subsection is devoted to
explicitly deriving these relations.

First we prove that for a purely magnetic impurity, the induced
(charge) LDOS is almost zero everywhere. We prove this by showing
$\text{Tr}[\delta{G}(\bq,\omega)]\approx0$. Within the
approximation, we
have\bea \nonumber \text{Tr}[\delta{G}(\bq,\omega)]&\approx&\int\frac{d^2k}{4\pi^2}\text{Tr}[G(\bk,\omega)\hat{V}G(\bk+\bq,\omega)]\\
\nonumber&=&V_0\int\frac{d^2k}{4\pi^2}\text{Tr}[G(\bk,\omega)\sigma_iG(\bk+\bq,\omega)].\\\eea
In the equation above we do not specify the spin polarization of the
impurity and the result is general. Noticing that the system is
invariant under time reversal operation ($C=i\sigma_y$), i.e.,
$CH(\bk)C^{-1}=H^T(-\bk)$ and that a magnetic impurity changes sign
under the same operation, i.e., $C\sigma_iC^{-1}=-\sigma_i^T$, we
have\bea \nonumber \nonumber
&&\int\frac{d^2k}{4\pi^2}\text{Tr}[G_0(\bk,\omega)\sigma_iG_0(\bk+\bq,\omega)]\\\nonumber
&=&\frac{d^2k}{4\pi^2}Tr[CG_0(\bk,\omega)\sigma_iG_0(\bk+bq,\omega)C^{-1}]\\\nonumber
&=&-\int\frac{d^2k}{4\pi^2}\text{Tr}[G_0^T(-\bk,\omega)\sigma_i^TG_0^T(-\bk-\bq,\omega)]\\\nonumber
&=&-\int\frac{d^2k}{4\pi^2}\text{Tr}[G_0(-\bk-\bq,\omega)\sigma_iG_0(-\bk,\omega)]\\
&=&0.\eea The last equality may be understood after changing
variables $\bk\rightarrow-\bk-\bq$.

Next we show that the approximate symmetries listed in
Table\ref{table1}(a) hold within the same approximation. According
to the table, we have, for impurity spin (again it is a purely
magnetic impurity) along the z-axis, the SLDOS
$S_z(x,y,t)\approx{S}_z(-x,y,t)$, which is equivalent to
$S_z(q_x,q_y,\omega)\approx{S}_z(-q_x,q_y,\omega)$. This may be
derived in the following way:
\begin{widetext}
\bea \nonumber
&&S_z(\bq,\omega)={\text{Tr}}[\delta{G}(\mathbf{q},\omega)\sigma_z]\approx\int\frac{d^2k}{(2\pi)^2}
\text{Tr}[G_0(\mathbf{k},\omega)\hat{V}G_0(\mathbf{k+q},\omega)\sigma_z]\\
\nonumber&=&V\int\frac{d^2k}{(2\pi)^2}\frac{\text{Tr}
[(\omega\sigma_0-k_y\sigma_x+k_x\sigma_y+\frac{\lambda}{2}
(k_+^3+k_-^3)\sigma_z)\sigma_z(\omega\sigma_0-(k_y+q_y)\sigma_x+(k_x+q_x)\sigma_y+\frac{\lambda}{2}((k+q)_+^3+(k+q)_-^3)\sigma_z)]}{((\omega+i\eta)^2-\epsilon_+^2(\mathbf{k}))((\omega+i\eta)^2-\epsilon_+^2(\mathbf{k+q}))}\\
\nonumber&=&V\int\frac{d^2k}{(2\pi)^2}(2\frac{\omega^2+k_y(k_y+q_y)+k_x(k_x+q_x)+\frac{\lambda^2}{4}(k_+^3+k_-^3)[(k+q)_+^3+(k-q)_-^3]}{((\omega+i\eta)^2-\epsilon_+^2(\mathbf{k}))((\omega+i\eta)^2-\epsilon_+^2(\mathbf{k+q}))})\\
&\approx&S_z(-q_x,q_y,\omega).\eea
\end{widetext}
In deriving the last equality, we notice that
$\epsilon(k_x,k_y)=\epsilon(-k_x,k_y)$ and change variables as
$k_x\rightarrow-k_x$. In the second column of Table\ref{table1}(a),
we find $S_z(x,y,t)\approx{S}_z(x,-y,t)$ for the impurity spin in x
and z directions. Since $\sigma_yH(k_x,k_y)\sigma_y=H(k_x,-k_y)$, we
have
\begin{widetext}
\bea
\nonumber S_z(q_x,q_y,\omega)&\approx&\int\frac{d^2k}{4\pi^2}\text{Tr}[G_0(k_x,k_y,\omega)\sigma_{x,z}G_0(k_x+q_x,k_y+q_y,\omega)\sigma_z]\\
\nonumber&=&\int\frac{d^2k}{4\pi^2}\text{Tr}[G_0(k_x,-k_y,\omega)(-\sigma_{x,z})G_0(k_x+q_x,-k_y-q_y,\omega)(-\sigma_z)]\\
\nonumber&=&\int\frac{d^2k}{4\pi^2}\text{Tr}[G_0(k_x,k_y,\omega)\sigma_{x,z}G_0(k_x+q_x,k_y-q_y,\omega)\sigma_z]\\
&=&S_z(q_x,-q_y,\omega).\eea
\end{widetext}
A simple consequence of the weak impurity approximation is a linear
combination of LDOS or SLDOS when there is more than one impurity,
or an impurity that has both magnetic and non-magnetic parts. In the
latter case, one can simply add up the FT-LDOS and FT-SLDOS for each
part to obtain the total configuration. But as discussed in the
text, the magnetic part contributes very little to the LDOS, and the
non-magnetic part does not contribute to the SLDOS (obvious from
time-reversal symmetry), most of the results for the magnetic
impurity part remain the same.

\section{Friedel oscillation at fixed energy in a 2D Dirac metal
by an edge impurity} In the text, we stated that when the energy
lies within the `Dirac regime' (e.g., when $\omega=0.05eV$), the
decay of the Friedel oscillation takes the form
$\rho(x,\omega)\propto{|x|}^{-3/2}$. In this subsection the
asymptotic expression for the LDOS oscillation caused by an edge
impurity in a 2D Dirac metal is derived. The Hamiltonian takes the
form\bea H(\bk)=v\bk\cdot\mathbf{\sigma}.\eea This form is
equivalent to the linear part in Eq(\ref{eq:model}) up to a global
spin-SU(2) gauge. The Hamiltonian may be easily solved: (only
positive energy
solutions are listed)\bea \epsilon(\bk)&=&vk,\\
\nonumber\psi(\bk)&=&(e^{i\phi/2},e^{-i\phi/2})^T/\sqrt{2},\eea
where $\phi$ being the polar angle. Now let us suppose that the
space is divided in half at $x=0$ (i.e., the edge impurity is
along y-axis), and the right side has a uniform potential of
$V=-V_0$ where $V_0>0$. The continuity of the
wavefunction at $x=0$ gives\bea \left(%
\begin{array}{c}
  e^{i\phi/2} \\
  e^{-i\phi/2} \\
\end{array}%
\right)+r(\phi)\left(%
\begin{array}{c}
  e^{i\phi'/2} \\
  e^{-i\phi'/2} \\
\end{array}%
\right)=t(\phi)\left(%
\begin{array}{c}
  e^{i\phi''/2} \\
  e^{-i\phi''/2} \\
\end{array}%
\right).\eea For the refraction part, $\phi''$ is fixed by
\bea\frac{vk+V_0}{v}\sin(\phi'')&=&k\sin(\phi).\eea And for the
reflection part, $\phi'=\pi-\phi$. Solving these equations, one
has \bea
r&=&-\frac{\sin(\frac{\phi-\phi''}{2})}{\cos(\frac{\phi+\phi''}{2})},\\
\nonumber t&=&\frac{\cos(\phi)}{\cos(\frac{\phi+\phi''}{2})}.\eea

Using the stationary phase approximation, we know that after
integrating all the $\bk$'s on the fixed energy contour
$\omega=vk$ to obtain the LDOS, the contribution mainly comes from
the $\bk$'s that have small polar angles. At small angles, the
reflection index takes the form\bea
r(\phi)\approx-\frac{V_0}{2(V_0+vk)}\phi.\eea Using
Eq(\ref{eq:edge_LDOS}), we
have\bea\nonumber\rho(x,\omega)&\approx&\int_{-\pi/2}^{\pi/2}\frac{d\phi}{2\pi}\frac{V_0}{V_0+\omega}\phi^2\sin(\frac{2\omega}{v}\cos(\phi)x)\\
\nonumber&\approx&\int_{-\infty}^{\infty}\frac{d\phi}{2\pi}\frac{V_0}{V_0+\omega}Im[\phi^2e^{i\frac{2\omega}{v}\cos(\phi)x}]\\
&\approx&\frac{V_0}{V_0+\omega}\frac{\sqrt{\pi}}{4}\cos(\frac{2\omega}{v}x-\frac{\pi}{4})(\frac{2\omega}{v}x)^{-3/2}.\eea

\end{document}